\documentclass[submission, Lecturenotes]{SciPost}
\usepackage{amsfonts}
\usepackage{amsmath}
\usepackage{amssymb}
\usepackage{amsthm}
\usepackage{amscd}

\usepackage{eucal}
\usepackage{bm}

\usepackage{booktabs}

\usepackage{graphicx}
\usepackage{float}

\input{macro.lib}

\allowdisplaybreaks

\begin{document}

\begin{center}{\Large \textbf{
Statistical mechanics of integrable quantum spin systems
}}\end{center}

% TODO: write the author list here. Use initials + surname format.
% Separate subsequent authors by a comma, omit comma at the end of the list.
% Mark the corresponding author with a superscript *.
\begin{center}
Frank G\"ohmann
\end{center}

\begin{center}
Fakult\"at f\"ur Mathematik und Naturwissenschaften\\
Bergische Universit\"at Wuppertal, Germany\\[1ex]
goehmann@uni-wuppertal.de
\end{center}

%\begin{center}
%{\LARGE \bf Statistical mechanics of integrable quantum spin systems}\\[7ex]
%{\large \sc Frank Göhmann}\\[2ex]
%Fakult\"at f\"ur Mathematik und Naturwissenschaften\\
%Bergische Universit\"at Wuppertal
%\end{center}

\vspace{10pt}
\noindent\rule{\textwidth}{1pt}
\tableofcontents\thispagestyle{fancy}
\noindent\rule{\textwidth}{1pt}
\vspace{10pt}

\clearpage
\section*{Preface}
This script is based on the notes the author prepared to give a set
of six lectures at the Les Houches School ``\emph{Integrability
in Atomic and Condensed Matter Physics}'' in the summer of 2018.
The responsibility for the selection of the material is partially
with the organisers, Jean-S\'ebastien Caux, Nikolai Kitanine,
Andreas Kl\"umper and Robert Konik. The school had its focus on the
application of integrability based methods to problems in non-equilibrium
statistical mechanics. My lectures were meant to complement this
subject with background material on the equilibrium statistical mechanics
of quantum spin chains from a vertex model perspective. I was asked
to provide a minimal introduction to quantum spin systems including
notions like the reduced density matrix and correlation functions
of local observables. I was further asked to explain the graphical
language of vertex models and to introduce the concepts of the
Trotter decomposition and the quantum transfer matrix. This was
basically the contents of the first four lectures presented at the
school. In the remaining two lectures I started filling these notions
with life by deriving an integral representation of the free energy
per lattice site for the Heisenberg-Ising chain (alias XXZ model)
using techniques based on non-linear integral equations.

Up to small corrections the following sections L1-L6 display the
lectures almost literally. The only major change is that the
example of the XXZ chain has been moved from section L5 to L2.
During the school it was not really necessary to introduce the model,
since other speakers had explained it before. But for these notes
I thought it might be useful to introduce the main example rather
early. I also supplemented each lecture with a comment section which
contains additional references and material of the type that was
discussed informally with the participants.

I am grateful to my colleagues at the University of Wuppertal,
Hermann Boos, Michael Karbach and Andreas Klümper, as well as to
my long-term collaborators Karol Kozlowski and Junji Suzuki for
sharing their considerable insight into the subjects of these
lectures. I would like to thank Constantin Babenko and Saskia
Faulmann for carefully reading the manuscript and pointing out a
number of misprints and inaccurracies in the first version.

\clearpage

\section{Statistical mechanics of quantum chains}
\subsection{Introduction}
Spin systems are the simplest conceivable quantum mechanical systems. In
nature the spin occurs in first place as an internal degree of freedom
of elementary particles. When many particles bind together in a many-body
quantum system like a crystalline solid, the spin may also take the
role of a discrete quantum number of collective excitations. In certain
experiments on such systems, e.g.\ on Mott insulators at low temperatures
or on ultra-cold atomic gases trapped in optical lattices, it is possible
to create `pure spin excitations'. Such systems are well described, in
a certain energy range, by (generalised) Hubbard or Heisenberg models
\cite{Thebook}, which are in the class of models to be considered in
these notes.

Spin systems can be used to illustrate the characteristic properties
of quantum mechanics, like its probabilistic nature or the superposition
and entanglement of states. For this reason quantum-spin systems are
ubiquitous in introductory text books on quantum mechanics \cite{FeynmanQM}
and familiar to all graduate students in physics.
\begin{itemize}
\item
A single spin-$\2$ is the simplest possible quantum system. Its
Hilbert space is ${\cal H} = {\mathbb C}^2$ equipped with the
Hermitian scalar product.
\item
Two spins-$\2$ describe the simplest interacting quantum systems
with  space of states ${\cal H}_2 = {\mathbb C}^2 \otimes {\mathbb C}^2$.
A two-spin state may be entangled.
\item
$N$ many interacting spins-$\2$ constitute the simplest many-body
quantum systems with space of states ${\cal H}_N = ({\mathbb C}^2)^{\otimes N}$.
Depending on the interaction, these systems may possess complicated highly
entangled ground states and may carry collective excitations of various
types.
\end{itemize}

In the thermodynamic (or `infinite volume') limit, $N \rightarrow + \infty$,
quantum-spin systems may exhibit critical behaviour. They can be used to
study phase transitions and quantum criticality. Apart from the number of
constituents $N$, quantum-spin systems typically depend on several interaction
parameters in their Hamiltonians. Considering certain scaling limits, in which
these parameters depend on $N$ and $N$ is send to infinity, quantum-spin
systems may be used to realise quantum field theories on the lattice.
Conversely, one may think of quantum-spin systems as of `fully regularised
quantum field theories' \cite{Faddeev95}, meaning that ultra-violet and infra-red
regularisations are provided by the fact that the number of spins is
finite, and the Hilbert space is regularised due to the fact that spins
have a finite number of degrees of freedom.

Arguably, all many-body quantum physics can be phrased in terms of spin
systems of sufficiently general type. This should provide enough motivation
to thoroughly study their statistical mechanics. It seems to indicate,
on the other hand, that the statistical mechanics of quantum-spin systems
in general should be too big as a subject for an introductory lecture
course. For this reason, after developing part of a general theory,
we shall restrict ourselves to integrable quantum-spin systems in
these notes. Integrable systems are defined in one spatial dimension (1d)
and have a rich algebraic structure underlying, which makes it possible
to obtain more or less explicit results, at least for the thermodynamics
of some non-trivial quantum spin systems in the infinite-volume limit.

As far as the general theory is concerned we shall introduce and
explain what we call the `quantum transfer matrix approach' to quantum
spin systems. In our understanding this approach is a clever and systematic
way of attaching an equilibrium statistical operator with any type of local
interaction. The equilibrium statistical operator of 1d spin systems will
be composed of transfer matrices of a `classical vertex model' on a
two-dimensional lattice. Such a procedure is non-unique. The non-uniqueness
may be utilised to optimise the properties of the transfer matrix for
various purposes, e.g.\ for an efficient calculation of the partition
function by quantum Monte Carlo methods \cite{Suzuki85}. We shall use
a construction, introduced by A. Kl\"umper in \cite{Kluemper93}, which
is optimised for the use with integrable quantum spin models of Yang-Baxter
type. As we shall see, this construction cannot only be used to calculate
the free energy per lattice site of such models, but seems to be optimised
as well for the calculation of their correlation functions \cite{GKS04a}.

\subsection{States and operators}
In our mathematical set-up we shall consider systems that are slightly
more general than spin-$\2$ systems in that they have $d \ge 2$
degrees of freedom on a `local Hilbert space' ${\cal H} = {\mathbb C}^d$,
equipped with the canonical Hermitian scalar product. We fix a basis
\begin{equation} \label{basish}
     \{e_\a\}_{\a = 1}^d \subset {\mathbb C}^d
\end{equation}
in this space.

\subsubsection{Operators on local Hilbert space}
In order to introduce a space of local observables we start with
a set of operators $e_\be^\a \in \End {\mathbb C}^d$, $\a, \be = 1,
\dots, d$, defined by their action on the basis~(\ref{basish}),
\begin{equation} \label{basisonbasish}
     e^\a_\be e_\g = \de^\a_\g e_\be \epp
\end{equation}

Then, for any $A \in \End {\mathbb C}^d$,
\begin{equation} \label{generalbasisaction}
     A e_\g = A^\be_\g e_\be = A^\be_\a \de^\a_\g e_\be = A^\be_\a e^\a_\be e_\g \epc
\end{equation}
where (\ref{basisonbasish}) was used in the last equation. In
(\ref{generalbasisaction}) we have also employed the common `summation
convention', implying that Greek indices that occur twice in an
expression are summed over from $1$ to $d$. We shall keep this
convention throughout these notes. Comparing left and right hand
sides of (\ref{generalbasisaction}) and taking onto account that
the set $\{e_\a\}_{\a = 1}^d$ is a basis, we conclude that
\begin{equation} \label{expandendh}
     A = A^\be_\a e^\a_\be
\end{equation}
and hence that
\begin{equation} \label{basisendh}
     \{e^\a_\be\}_{\a, \be = 1}^d \subset \End {\mathbb C}^d
\end{equation}
is a basis of $\End {\mathbb C}^d$. Any basis element $e^\a_\be$
will be called an elementary endomorphism.

The action of a product of two elementary endomorphisms on the basis
(\ref{basish}) can be computed by means of (\ref{generalbasisaction}),
\begin{equation}
     e^\a_\be e^\g_\de e_\ph = e^\a_\be \de^\g_\ph e_\de
        = \de^\a_\de \de^\g_\ph e_\be = \de^\a_\de e^\g_\be e_\ph \epp
\end{equation}
Comparing the left and the right hand side of this equation and
using again that $\{e_\a\}_{\a = 1}^d$ is a basis of ${\mathbb C}^d$,
we obtain the relation
\begin{equation}
     e^\a_\be e^\g_\de= \de^\a_\de e^\g_\be
\end{equation}
providing a set of structure constants for the algebra $\End
{\mathbb C}^d$.

From the first equation (\ref{generalbasisaction}) we see that the
identity operator $I_d \in \End {\mathbb C}^d$ is represented by the
matrix $A_\a^\be = \de_\a^\be$. Hence, by (\ref{expandendh}),
\begin{equation}
     I_d = e^\a_\a \epp
\end{equation}

\subsubsection{\boldmath Local basis of $L$-site Hilbert space}
A multi-spin system is defined on a regular or irregular lattice
consisting of $N$ points $\xv_k \in {\mathbb R}^n$, $k = 1, \dots, N$,
such that a local Hilbert space ${\cal H} = {\mathbb C}^d$ is associated
with every point. The Hilbert space of the multi-spin system is
then ${\cal H}_N = ({\mathbb C}^d)^{\otimes N}$. Since we will soon
focus on large integrable lattice systems, we shall assume that
$n = 1$ and $\xv_k = - L + k$, $k = 1, \dots, N = 2L$.

We define the embedding of the basis of elementary endomorphisms
(\ref{basisendh}) into the lattice,
\begin{equation}
     {e_j}_\a^\be = I_d^{\otimes (L - 1 + j)} \otimes e_\a^\be
                    \otimes I_d^{\otimes (L - j)}
		    \in \End ({\mathbb C}^d)^{\otimes 2L}
\end{equation}
for $j = - L + 1, \dots, L$. With this we can embed the action
of `$m$-site operators' into the lattice. For every $A \in
\End ({\mathbb C}^d)^{\otimes m}$, $m \le 2L$, and $\{j_1, \dots,
j_m\} \subset \{- L + 1, \dots, L\}$ we set
\begin{equation}
     A_{j_1 \dots j_m} = A^{\a_1 \dots \a_m}_{\be_1 \dots \be_m}
                         {e_{j_1}}_{\a_1}^{\be_1} \dots {e_{j_m}}_{\a_m}^{\be_m} \epp
\end{equation}

\subsubsection{Examples}
\begin{enumerate}
\item
If $A \in \End {\mathbb C}^d$, then $A_j$ is a `single-site' (or
ultra-local) operator acting on `site $j$'.
\item
Let
\begin{equation} \label{defp}
     P = e_\a^\be \otimes e_\be^\a \in \End {\mathbb C}^d \otimes {\mathbb C}^d \epp
\end{equation}
Then
\begin{multline}
     P \, \xv \otimes \yv = e_\a^\be \otimes e_\be^\a \; x^\g e_\g \otimes y^\de e_\de
        = x^\g y^\de \; e_\a^\be e_\g \otimes e_\be^\a e_\de \\
	= x^\g y^\de \de^\be_\g \de^\a_\de \; e_\a \otimes e_\be = \yv \otimes \xv \epp
\end{multline}
Thus, $P$ induces the transposition of factors in a tensor product.
In physical terms, it interchanges the states on two sites. This
operator, called the transposition or exchange operator, is an
important object in the theory of spin systems and occurs in many
places.

Most prominently, perhaps, it occurs as `exchange interaction' in the
Heisenberg Hamiltonian
\begin{equation} \label{gldham}
    H = J \sum_{j = - L + 1}^L P_{j-1,j} \epc
\end{equation}
where $J > 0$ and $P_{- L, - L + 1} = P_{L, - L + 1}$ by definition.
This Hamiltonian defines one of the simplest and most generic
interacting quantum-spin systems. It is simple in the sense that only
neighbouring sites interact and also because $P (A \otimes A) =
(A \otimes A) P$ implies that
\begin{equation}
     (A \otimes A)  P (A^{-1} \otimes A^{-1}) = P
\end{equation}
for all $A \in GL (d)$, the general linear group of invertible
endomorphisms on ${\mathbb C}^d$. This includes, in particular, the
case, when $A$ is a $d$-dimensional representation of the group
of rotations $SO(3)$. The Hamiltonian is sometimes called the
`$GL(d)$-invariant magnet'. In the literature the term `Heisenberg
model' is often reserved for the case $d = 2$.

The operator $P$ also plays in important role for the implementation
of the action of spatial symmetries on quantum-spin systems, since
it induces the action of the symmetric group ${\mathfrak S}^{2L}$ on
${\mathcal H}_{2L}$. For $j, k, l \in \{-L+1, \dots, L\}$ mutually
distinct
\begin{subequations}
\begin{align}
     & P_{jk} {e_k}_\a^\be = {e_j}_\a^\be P_{jk} \epc \\
     & P_{jk}^2 = \id \epc \label{idem} \\
     & P_{jk} P_{kl} = P_{jl} P_{jk} = P_{kl} P_{jl} \label{exchangep} \epc
\end{align}
\end{subequations}
which follows immediately from the definition (\ref{defp}) of $P$.
The braid relation
\begin{equation}
     P_{j j+1} P_{j+1 j+2} P_{j j+1} = P_{j+1 j+2} P_{j j+1} P_{j+1 j+2}
\end{equation}
follows from (\ref{idem}), (\ref{exchangep}). Braid relation and
(\ref{idem}) define the symmetric group.

The symmetry group of a spin chain with an even number of sites is
the dihedral group ${\cal D}_{2L} = {\cal C}_{2L} \rtimes {\cal C}_2
\subset {\mathfrak S}^{2L}$ which is the symmetry group of a regular
polygon with $2L$ edges. Being a product of two cyclic groups it has
two generators
\begin{subequations}
\begin{align}
     & \hat U = P_{-L+1, -L+2} \dots P_{L-1, L} \epc \\
     & \hat P = P_{-L+1, L} P_{-L+2, L-1} \dots P_{0,1} \epc
\end{align}
\end{subequations}
the `shift operator' $\hat U$ and the `parity operator' $\hat P$.
Note that $\hat U^{2L} = \hat P^2 = \id$.

Here is a third example for the occurance of $P$ in the theory of
quantum-spin systems. It provides a family of rational (or `Yangian')
solutions of the Yang-Baxter equation which is behind the
integrability of the $GL(d)$ invariant Hamiltonians (\ref{gldham}).
Define
\begin{equation} \label{rrational}
     R(\la, \m) = (\la - \m) I_d \otimes I_d + P \epp
\end{equation}
Then, using (\ref{exchangep}), it is easy to see that
\begin{equation}
     R_{jk} (\la, \m) R_{jl} (\la, \nu) R_{kl} (\m, \nu) =
        R_{kl} (\m, \nu) R_{jl} (\la, \nu) R_{jk} (\la, \m) \epc
\end{equation}
if $j, k , l \in \{-L+1, \dots, L\}$ are mutually distinct.
\end{enumerate}

\subsection{Interactions}
In the following we shall focus on quantum-spin chains with
Hilbert space ${\cal H}_{2L}$ and with local interactions $h \in \End
({\mathbb C}^d)^{\otimes m}$, where $m \in \{2, \dots, 2L\}$ will
be called the range of the interaction. Setting
\begin{equation}
     h_{j, j+1, \dots, j+m-1} = \hat U^{j-1} h_{1, \dots, m} \hat U^{1-j}
\end{equation}
for $j = -L+1, \dots, L$ we define the Hamiltonian
\begin{equation}
     H = \sum_{j = -L+1}^L h_{j, j+1, \dots, j+m-1}
\end{equation}
which is translation invariant (`satisfies periodic boundary conditions')
by construction.

\subsection{Statistical mechanics of quantum-spin systems}
It is part of the wisdom of quantum statistical mechanics that `large
quantum systems' cannot be prepared in a pure quantum state
as they cannot be fully separated from `the observer'. After
preparation in an experiment any large quantum-spin system will rather
be in a state described by a density matrix (a statistical operator)
$\r_L \in \End {\cal H}_{2L}$ with properties
\begin{equation}
     \r_L = \r_L^+ \epc \qd \r_L \ge 0 \epc \qd \tr \r_L = 1 \epp
\end{equation}
These properties guarantee that $\r_L$ is diagonalizable and that
the spectrum of $\r_L$ is a discrete probability distribution. We
may think of $\r_L$ as representing an ensemble. Subsequent
experiments then measure ensemble averages
\begin{equation}
     \<X\> = \tr_{-L+1, \dots, L} \{\r_L X\}
\end{equation}
of operators $X \in \End_{{\cal H}_{2L}}$.

In general $\r_L$ is time-dependent and its time dependence (in
the Schr\"odinger picture!) is determined by the von-Neumann
equation
\begin{equation}
     \i \6_t \r_L = [H, \r_L] \epp
\end{equation}
Hence, a stationary density matrix should be a function of the
conserved quantities commuting with the Hamiltonian.
\subsubsection{Examples}
In these notes we restrict ourselves to stationary density matrices.
Some important examples are listed below.
\begin{enumerate}
\item
Many-body quantum systems cannot be separated from their environment
forever. Eventually they relax to the canonical ensemble,
\begin{equation} \label{candens}
     \r_L (T) = \frac{\re^{- H/T}}{\tr_{-L+1, \dots, L} \{\re^{- H/T}\}} \epp
\end{equation}
However, transients and long relaxation times are possible, particularly
for integrable quantum-spin systems, and stationary non-equilibrium
ensembles may be realised in driven systems.
\item
Two more examples are the zero and infinite-temperature limits of $\r_L (T)$.
\begin{equation}
     \lim_{T \rightarrow 0+} \r_L (T) = \frac{1}{g} \sum_{j=1}^g |\ps_j^{(0)}\>\<\ps_j^{(0)}| \epc
\end{equation}
where the $\{|\ps_j^{(0)}\>\}_{j=1}^g$ form an orthonormal basis of the
ground-state sector of ${\cal H}_{2L}$ and $g$ is the ground-state
degeneracy, and
\begin{equation}
     \lim_{T \rightarrow + \infty} \r_L (T) = d^{-2L} \cdot \id \epp
\end{equation}
\item
A special case of ensembles are those represented by any excited state
$|\ps_n\>$ of the Hamiltonian $H$. The corresponding density matrices are
\begin{equation}
     \r_L^{(n)} = |\ps_n\>\<\ps_n| \epp
\end{equation}
\end{enumerate}
\subsubsection{Integrable quantum-spin systems}
These notes will focus on large integrable quantum-spin systems. The problems
we are going to address comprise:
\begin{enumerate}
\item
A description of density matrices $\r_L$ in a way compatible with
the integrable structure,
\item
the calculation of the free energy per lattice site in the thermodynamic
limit,
\begin{equation}
     f(T) = - T \lim_{L \rightarrow + \infty} \frac{1}{L} \ln \bigl(
                \tr_{-L+1, \dots, L} \{\re^{- H/T}\} \bigr) \epc
\end{equation}
\item
the calculation of two-point functions of local operators
in the thermodynamic limit,
\begin{equation}
     \<x_1 y_{m+1}\> = \lim_{L \rightarrow + \infty} \tr_{-L+1, \dots, L}
                       \bigl\{ \r_L (T) x_1 y_{m+1} \bigr\} \epp
\end{equation}
\end{enumerate}
\subsection{Comments}
It is an interesting problem to describe the class of
density matrices that may be generated as a result of the
relaxation of an isolated integrable quantum spin system
towards equilibrium. Due to the existence of a large number of
additional local conserved quantities, the class of such
density matrices must be much larger than the one-parametric
family (\ref{candens}). In \cite{RDYO07} the concept of generalised
Gibbs ensembles was suggested with an inverse-temperature like
Lagrange parameter for every additional conserved quantity.
Conceptual difficulties arise from the fact that it is hard
to identify `a complete set of conserved local operators'
in an infinite integrable system. We recommend \cite{IDWCEP15,%
EsFa16,ViRi16} for further reading.

\section{The quantum transfer matrix}
\subsection{Fundamental models}
A sufficiently general setting for a statistical mechanics of
integrable quantum-spin systems is that of `fundamental Yang-Baxter
integrable models'. Fundamental integrable spin systems are entirely
defined in terms of a matrix $R(\la,\m): {\mathbb C}^2 \mapsto
\End \bigl( {\mathbb C}^d \otimes {\mathbb C}^d \bigr)$ which satisfies
\begin{subequations}
\label{funda}
\begin{align} \label{ybe}
     & R_{12} (\la, \m) R_{13} (\la,\nu) R_{23} (\m, \nu) =
          R_{23} (\m, \nu) R_{13} (\la,\nu) R_{12} (\la, \m) \epc \\[1ex]
     & R(\la,\la) = P \epp \label{reg}
\end{align}
\end{subequations}
Equation (\ref{ybe}) is the famous Yang-Baxter equation. It is
this equation that underlies the integrability of many quantum-spin
systems. Solutions of the Yang-Baxter equation are called $R$-matrices.
An $R$-matrix satisfying equation (\ref{reg}) is called regular.
The arguments of $R(\la ,\m)$ are called spectral parameters.

Assuming differentiability in $\la, \m$ in a vicinity of $(0,0)$
equations (\ref{funda}) imply another property of the $R$-matrix
which is called unitarity: There is a function $g: {\mathbb C}^2
\mapsto {\mathbb C}$, differentiable in a neighbourhood of $(0,0)$,
$g(0,0) = 1$, $g(\la, \m) = g(\m, \la)$, such that
\begin{equation} \label{unitary}
     \frac{R_{12} (\la, \m) R_{21} (\m, \la)}{g(\la, \m) g(\m, \la)} = \id \epp
\end{equation}
We may therefore assume in the following that $R$ is normalised in
such a way that
\begin{equation} \label{uni}
     R_{12} (\la, \m) R_{21} (\m, \la) = \id \epp
\end{equation}
The proof of the existence of the function $g$ is left as an exercise
to the reader. 

With any $R$-matrix satisfying (\ref{funda}), (\ref{uni}) we 
associate two transfer matrices
\begin{subequations}
\begin{align}
     & t_\perp (\la) = \tr_a \{R_{a,L} (\la, 0) \dots R_{a, -L+1} (\la, 0)\} \epc \\[1ex]
     & \overline{t}_\perp (\la) = \tr_a \{R_{-L+1,a} (0,\la) \dots R_{L,a} (0,\la)\}
\end{align}
\end{subequations}
and a Hamiltonian
\begin{equation} \label{hamfun}
     H = h_R t_\perp' (0) t_\perp^{-1} (0)
         = - h_R t_\perp (0) \overline{t}_\perp' (0)
         = h_R \sum_{j = -L+1}^L \6_\la (P R)_{j-1, j} (\la, 0)\bigr|_{\la = 0} \epc
\end{equation}
where $(PR)_{-L, -L+1} = (PR)_{L, -L +1}$ by definition and where $h_R
\in {\mathbb C}$ is a constant which may be used to render $H$
Hermitian and to set the energy scale (alternatively one may rescale
the spectral parameter). In order to obtain the second equation in
(\ref{hamfun}) one has to use (\ref{uni}).

\subsection{Trotter formula}
Let
\begin{equation}
     X_N = \frac{t_\perp \bigl(- \frac{h_R}{2NT}\bigr)
                 \overline{t}_\perp \bigl(\frac{h_R}{2NT}\bigr) - 1}{1/N}
\end{equation}
and observe that, due to (\ref{hamfun}),
\begin{equation}
     \lim_{N \rightarrow + \infty} X_N = - \frac{H}{T} \epp
\end{equation}
It is not difficult to see that
\begin{equation}
     \Bigl\| \re^{- H/T} - \Bigl[ t_\perp \Bigl(- \frac{h_R}{2NT}\Bigr)
                             \overline{t}_\perp \Bigl(\frac{h_R}{2NT}\Bigr)\Bigr]^N \Bigr\| \le
     \Bigl\| \re^{- H/T} - \re^{X_N} \Bigr\| + \frac{\|X_N\|^2}{2N} \re^{\|X_N\|} \epc
\end{equation}
where $\| \cdot \|$ is the operator norm. Hence,
\begin{equation} \label{trotter}
     \re^{- H/T} =  \lim_{N \rightarrow + \infty} 
                      \Bigl[ t_\perp \Bigl(- \frac{h_R}{2NT}\Bigr)
		      \overline{t}_\perp \Bigl(\frac{h_R}{2NT}\Bigr)\Bigr]^N \epp
\end{equation}
This way the (unnormalised) density matrix of the canonical ensemble 
is represented as a product of transfer matrices. Equation (\ref{trotter})
is sometimes called `the Trotter formula', $N$ `the Trotter number'.
\subsection{External fields}
Assume there is $\Th (\a) = \re^{\a \hat \ph}$ with $\hat \ph \in
\End ({\mathbb C}^d)$, $\a \in {\mathbb C}$ such that
\begin{equation} \label{u1sym}
     [R_{12} (\la,\m), \Th_1 (\a) \Th_2 (\a)] = 0 \epc
\end{equation}
which is called a $U(1)$ symmetry of the $R$-matrix. Then
\begin{equation}
     [t_\perp (\la), \Th_{-L+1} (\a) \dots \Th_L (\a)] =
        [\, \overline{t}_\perp (\la), \Th_{-L+1} (\a) \dots \Th_L (\a)] = 0 \epp
\end{equation}
Setting
\begin{equation}
     \hat \PH = \sum_{j = - L + 1}^{L} \hat \ph_j
\end{equation}
we conclude that
\begin{equation}
     [t_\perp (\la), \hat \PH] = [\, \overline{t}_\perp (\la), \hat \PH] = 0 \epp
\end{equation}

Setting
\begin{equation} \label{hl}
     H_L = H - \k \hat \PH
\end{equation}
this allows us to couple an external field to the Hamiltonian
without spoiling its integrability.
\subsection{Quantum transfer matrix}
For $N$ even introduce `vertical spaces' $\bar 1, \dots, \overline N$.
By definition
\begin{equation} \label{stimonodromy}
     T_a (\la|\a) = \Th_a (\a) R^{t_1}_{\overline N, a} (\n_N, \la)
                    R_{a, \overline{N-1}} (\la, \n_{N-1}) \dots
		    R^{t_1}_{\bar 2, a} (\n_2, \la)
                    R_{a, \bar 1} (\la, \n_1)
\end{equation}
is the `staggered and twisted inhomogeneous monodromy matrix' of the
fundamental model. Here
\begin{equation}
     R^{t_1} (\la, \m) = R^{\a \g}_{\be \de} e^\a_\be \otimes e^\de_\g \epp
\end{equation}
The Yang-Baxter equation (\ref{ybe}) implies
\begin{equation} \label{yba}
     R_{ab} (\la, \m) T_a (\la|\a) T_b (\m|\a)
        = T_b (\m|\a) T_a (\la|\a) R_{ab} (\la, \m)
\end{equation}
(Exercise: Prove it! Hint: first show that (\ref{ybe}) implies $R_{ab} (\la,\m)
R_{j, a}^{t_1} (\nu, \la) R_{jb}^{t_1} (\n, \m) = R_{jb}^{t_1} (\n, \m) 
R_{j, a}^{t_1} (\nu, \la) R_{ab} (\la,\m)$). These are the `Yang-Baxter
algebra relations' for the (operator-valued) matrix elements of
$T_a (\la|\a)$ considered as a matrix in `auxiliary space $a$'.

We shall call the transfer matrix associated with $T_a (\la|\a)$ the
`quantum transfer matrix' of the fundamental model and denote it by
\begin{equation}
     t(\la|\a) = \tr_a \bigl\{T_a (\la|\a)\bigr\} \epp
\end{equation}
Using (\ref{trotter}) and the Yang-Baxter equation (\ref{ybe}) 
we obtain
\begin{multline} \label{rhoqtm}
     \re^{- H_L/T} = \re^{\k \hat \PH/T} \re^{- H/T} =
        \lim_{N \rightarrow \infty} \re^{\k \hat \PH/T}
                      \Bigl[ t_\perp \Bigl(- \frac{h_R}{NT}\Bigr)
		      \overline{t}_\perp \Bigl(\frac{h_R}{NT}\Bigr)\Bigr]^\frac{N}{2} \\[1ex]
     = \lim_{N \rightarrow \infty}
       \tr_{\bar 1, \dots, \overline N} \bigr\{ T_{- L + 1} (0) \dots T_L (0) \bigr\}
\end{multline}
if we set $\n_{2j-1} = h_R/(NT)$, $\n_{2j} = - h_R/(NT)$, $j = 1, \dots, N/2$,
and $\a = \k/T$.

It is instructive (and not too hard) to derive (\ref{rhoqtm}) by algebraic
means (see e.g.\ \cite{GKS04a}). In the following, however, we shall
introduce the graphical language of vertex models and use it for an
intuitive and easily memorizable proof.
\subsection{Example: the XXZ chain}
The basic example of a fundamental integrable model is the XXZ
spin-$\2$ chain. Its $R$-matrix (the $R$-matrix of the six-vertex
model \cite{Babook}) can be understood as a `$q$-deformation' of
the rational $R$-matrix (\ref{rrational}) for $d = 2$. With a
rescaling appropriate for our purposes it becomes
\begin{subequations}
\label{rxxz}
\begin{align}
     R(\la,\m) & = \begin{pmatrix}
                    1 & 0 & 0 & 0 \\
		    0 & b(\la - \m) & c(\la - \m) & 0 \\
		    0 & c(\la - \m) & b(\la - \m) & 0 \\
		    0 & 0 & 0 & 1
		   \end{pmatrix} \epc
		   \displaybreak[0] \\[2ex] \label{defbc}
     b(\la) & = \frac{\sh(\la)}{\sh(\la + \h)} \epc \qd
     c(\la) = \frac{\sh(\h)}{\sh(\la + \h)} \epp
\end{align}
\end{subequations}
It is a simple exercise to verify that (\ref{rxxz}) describes
a one-parameter family of solutions of the Yang-Baxter
equation (\ref{ybe}) that is regular (\ref{reg})
and unitary (\ref{uni}).

In order to generate the Hamiltonian we differentiate
\begin{multline}
     \6_\la P R(\la, 0)\bigr|_{\la = 0} =
        \begin{pmatrix}
	   0 &  &  &  \\ & c' (0) & b' (0) & \\
	   & b' (0) & c' (0) &  \\ & & & 0
	\end{pmatrix} = \frac{1}{\sh (\h)}
        \begin{pmatrix}
	   0 &  &  &  \\ & - \D & 1 & \\
	   & 1 & - \D &  \\ & & & 0
	\end{pmatrix} \\[1ex]
	= \frac{1}{2 \sh (\h)}
	  \bigl\{ \s^x \otimes \s^x + \s^y \otimes \s^y
	          + \D \bigl(\s^z \otimes \s^z - \id \bigr) \bigr\} \epc
\end{multline}
where $\s^\a$, $\a = x, y, z$, are Pauli matrices and $\D = \ch(\h)$
by definition. Setting $h_R = 2 J \sh(\h)$ in our general formula
(\ref{hamfun}) we obtain the XXZ Hamiltonian
\begin{equation} \label{hxxz}
     H_{XXZ} = J \sum_{j = - L + 1}^L \bigl\{ \s_{j-1}^x \s_j^x + \s_{j-1}^y \s_j^y
                 + \D \bigl( \s_{j-1}^z \s_j^z - 1 \bigr) \bigr\} \epc
\end{equation}
where $\s_{-L}^\a = \s_L^\a$ by definition.  $H_{XXZ}$ is hermitian
for all real $J$ and $\D$. A closer inspection of its discrete
symmetries reveals that we may restrict ourselves to $J > 0$.

Clearly
\begin{equation}
     [R(\la, \m), \re^{\a \s^z/2} \otimes \re^{\a \s^z/2}] = 0 \epc
\end{equation}
meaning that $R(\la, \m)$ has a $U(1)$-symmetry generated by
$\Th (\a) = \re^{\a \hat \ph}$ with $\hat \ph = \s^z/2$.
Thus, in this case
\begin{equation}
     H_L = H_{XXZ} - \frac{\k}{2} \sum_{j = - L + 1}^L \s_j^z
\end{equation}
(cf.\ (\ref{hl})) and $\k$ has the meaning of a Zeeman magnetic
field coupling to the individual spins.

\subsection{Graphical representation of integrability objects}
A rather efficient way of dealing with the relation between the
various `integrability objects' introduced above, the $R$-matrix
and the transfer and monodromy matrices, utilises a certain graphical
representation \cite{Babook}.
\begin{enumerate}
\item
We identify every matrix element of $R(\la, \m)$ with a `vertex',
\begin{equation}
     R^{\a \g}_{\be \de} (\la, \m) \: = \:
        \text{\raisebox{-58pt}{\includegraphics[width=.26\textwidth]{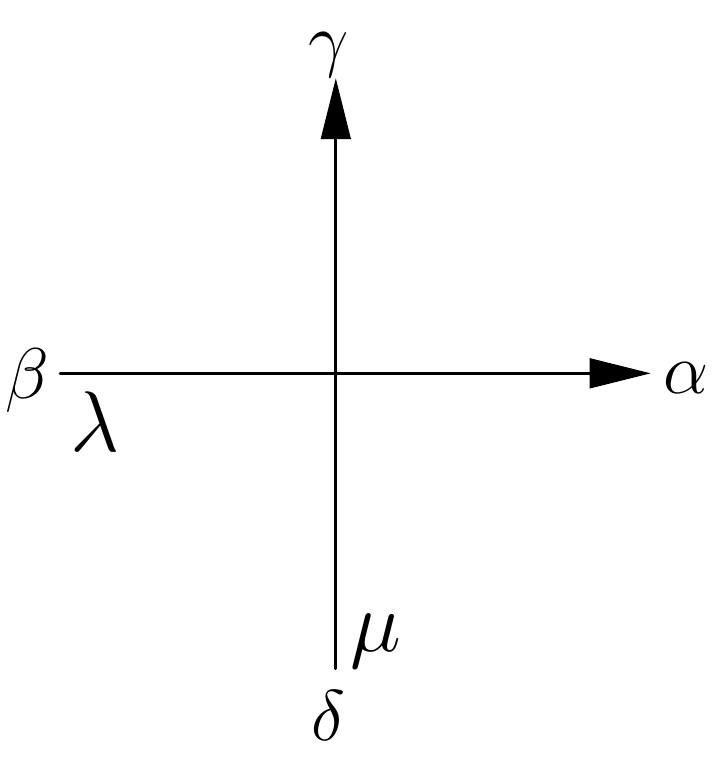}}} \epc
\end{equation}
$\a, \be, \g, \de = 1, \dots, d$ (for $d = 2$ also $\pm$ or $\auf, \ab$).
\item
Every arrangement of a finite number of directed, crossing lines is
then in one-to-one correspondence with a product of $R$-matrix
elements. We identify the connection of lines with summation over indices,
e.g.
\begin{equation}
     \text{\raisebox{-50pt}{\includegraphics[width=.50\textwidth]{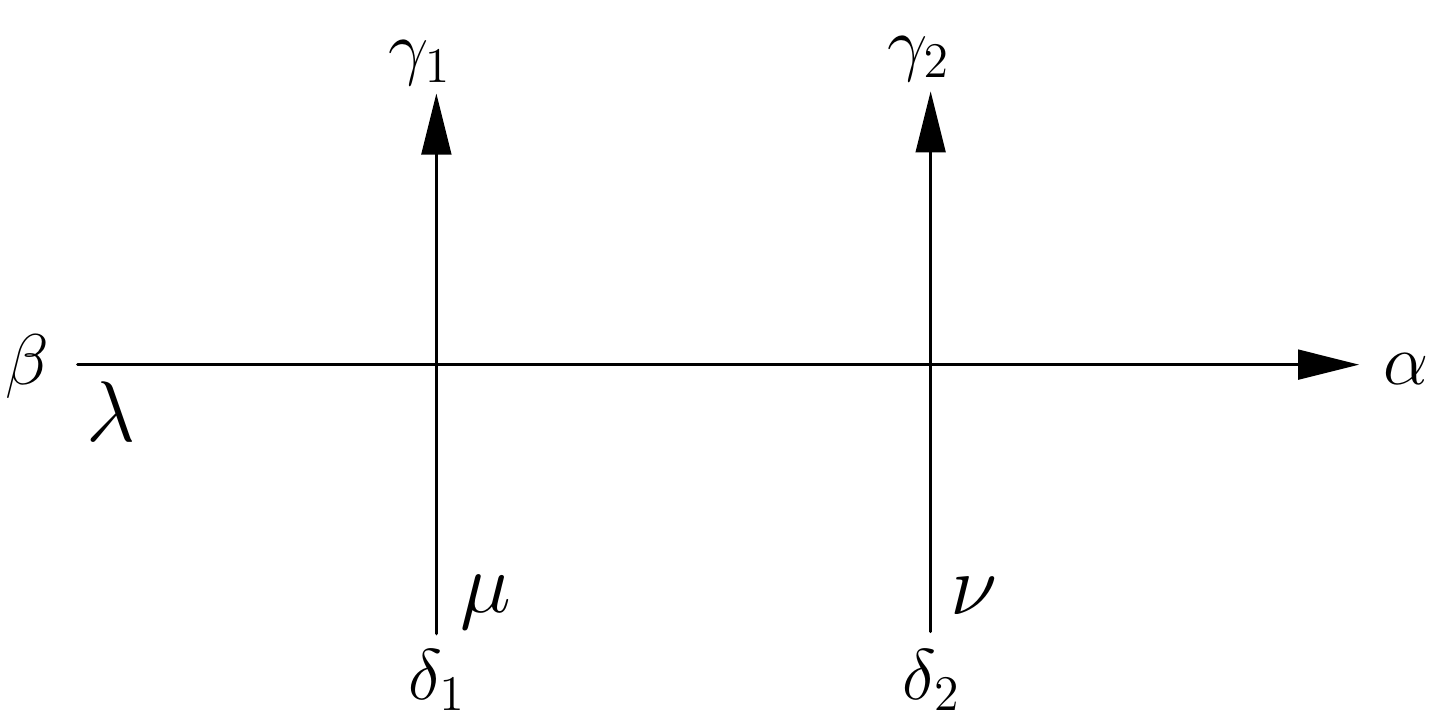}}}
	\: = \: R^{\a \g_2}_{\be' \de_2} (\la, \n) R^{\be' \g_1}_{\be \de_1} (\la, \m) \epp
\end{equation}
\item
We indicate closed lines by a small semi-loop at the tail, e.g.
\begin{equation}
     \text{\raisebox{-36pt}{\includegraphics[width=.24\textwidth]{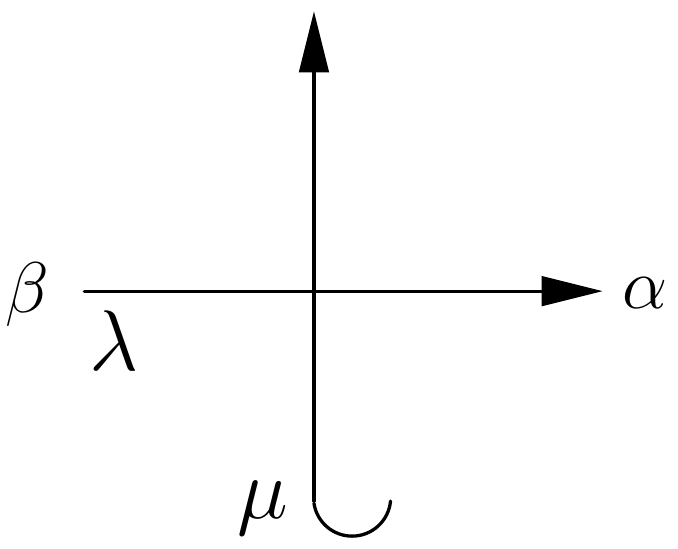}}}
        \: = \: R^{\a \g}_{\be \g} (\la, \m) \: = \:
        \text{\raisebox{-41pt}{\includegraphics[width=.30\textwidth]{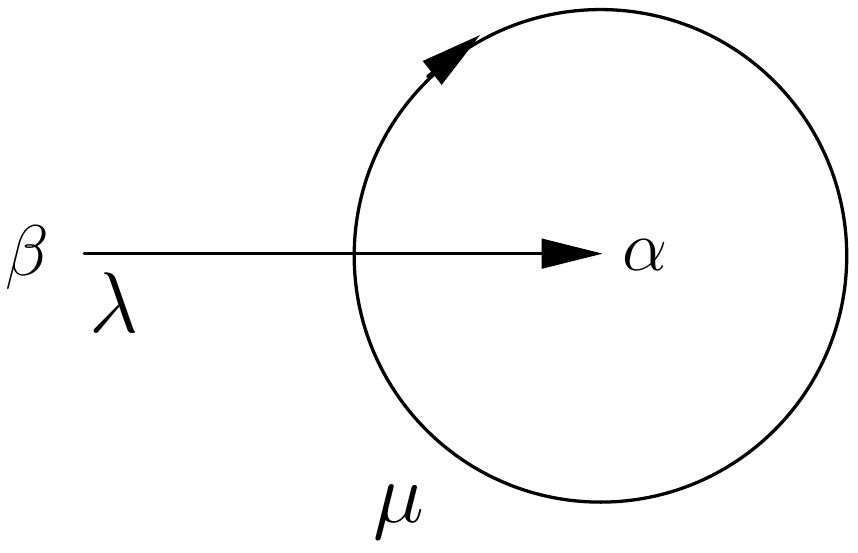}}} \epp
\end{equation}
\item
This way we obtain a graphical representation of the Yang-Baxter
equation:
\begin{equation} \label{graphybe}
\includegraphics[width=.86\textwidth]{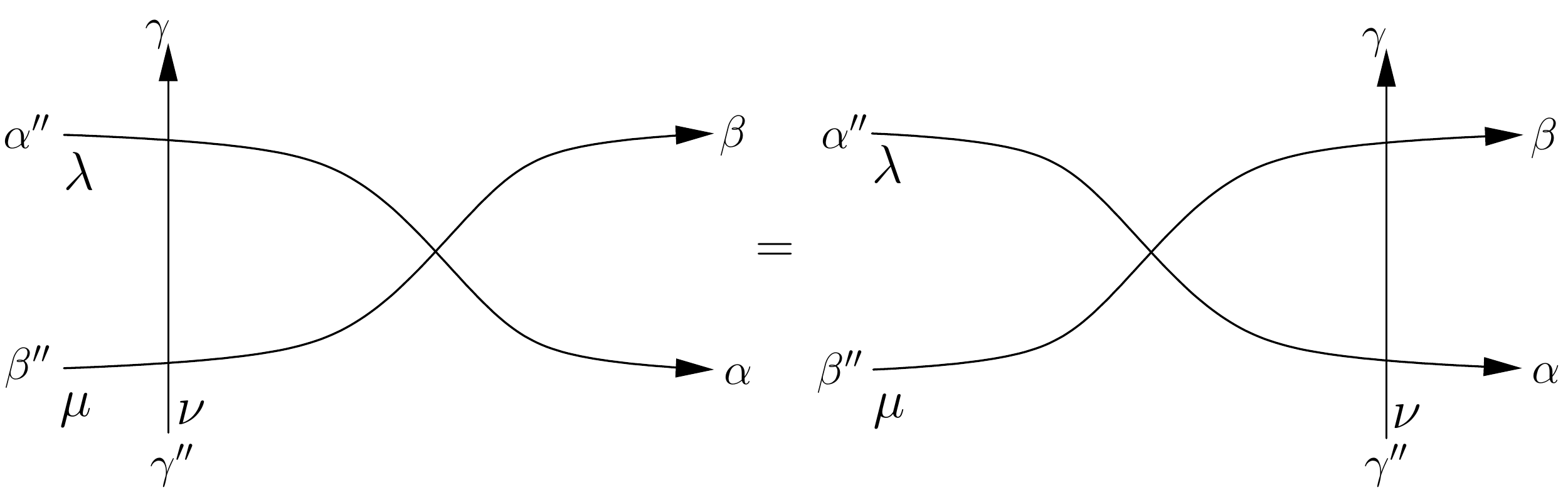}
\end{equation}
translates into
\begin{equation}
     R^{\a \be}_{\a' \be'} (\la, \m) R^{\a' \g}_{\a'' \g'} (\la, \n)
        R^{\be' \g'}_{\be'' \g''} (\m, \n) = R^{\be \g}_{\be' \g'} (\m, \n)
	R^{\a \g'}_{\a' \g''} (\la, \n) R^{\a' \be'}_{\a'' \be''} (\la, \m)
\end{equation}
which is the coordinate form of the Yang-Baxter equation (\ref{ybe})
(Exercise: Verify!).
\item
For consistency we need the rule
\begin{equation}
     \text{\raisebox{-15pt}{\includegraphics[width=.30\textwidth]{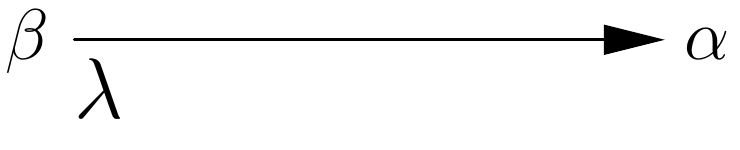}}}
        \: = \: \de^\a_\be \epp
\end{equation}
\item
Then regularity (\ref{reg}) has the graphical representation
\begin{equation} \label{graphreg}
     \text{\raisebox{-42pt}{\includegraphics[width=.55\textwidth]{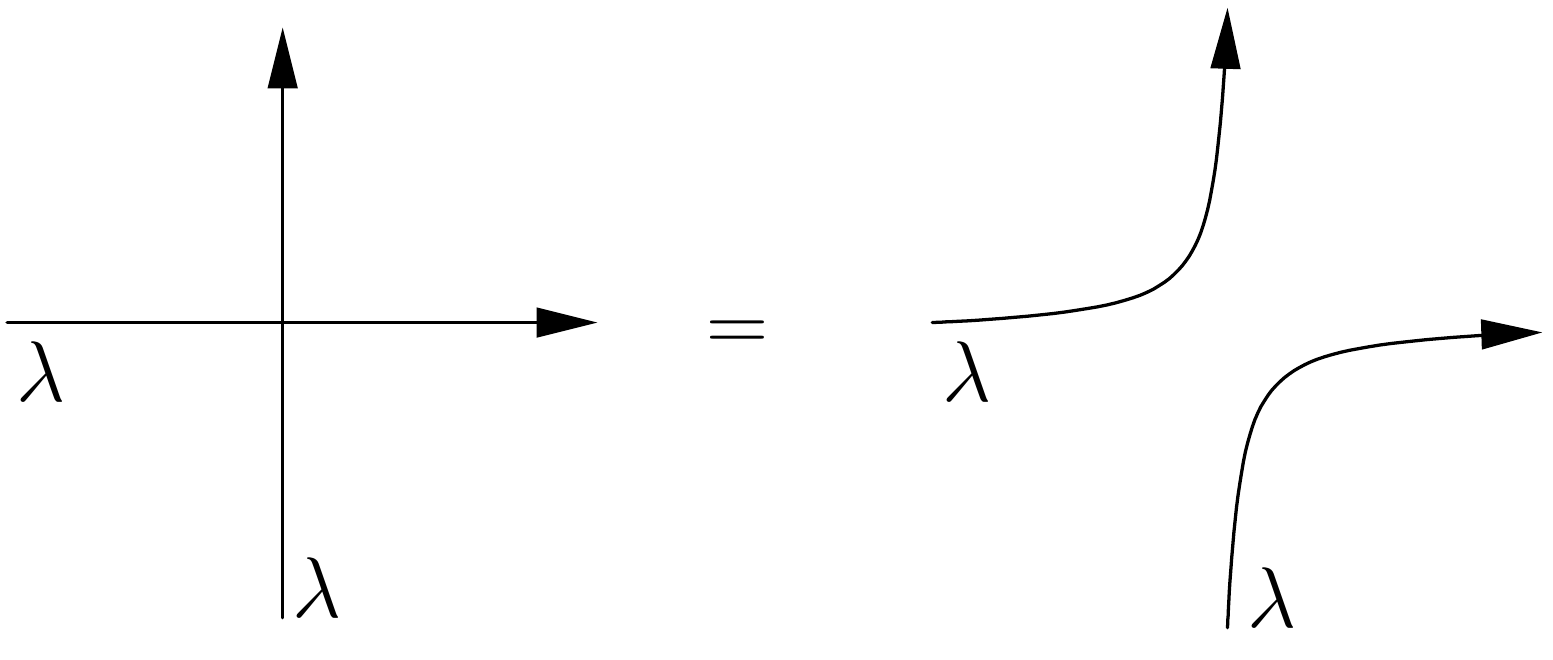}}} \epp
\end{equation}
\item
And unitarity is drawn as
\begin{equation}
     \text{\raisebox{-28pt}{\includegraphics[width=.83\textwidth]{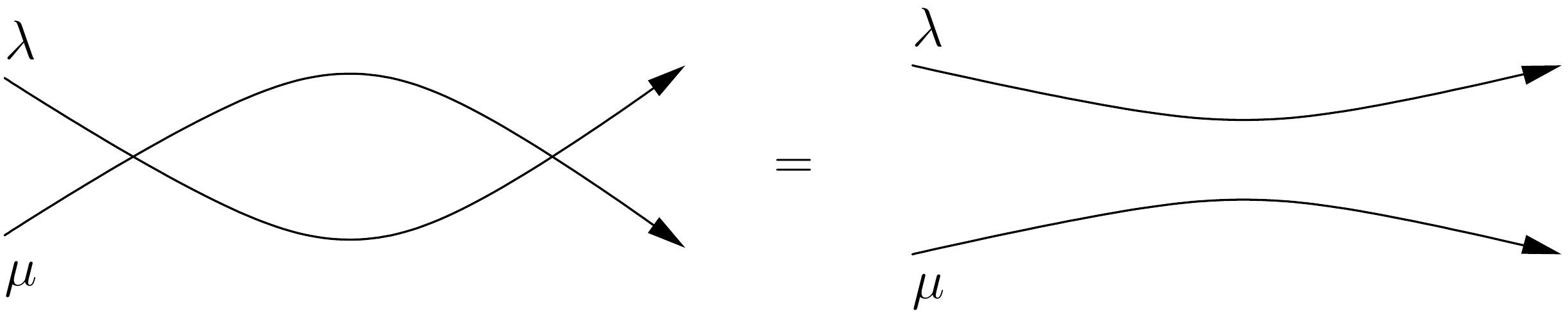}}} \epp
\end{equation}
(Exercise: Show (vi) and (vii)!)
\item
Single-site operators, such as $\Th (\k)$ can be represented as
\begin{equation}
     \Th^\a_\be (\k) \: = \:
        \text{\raisebox{-14pt}{\includegraphics[width=.24\textwidth]{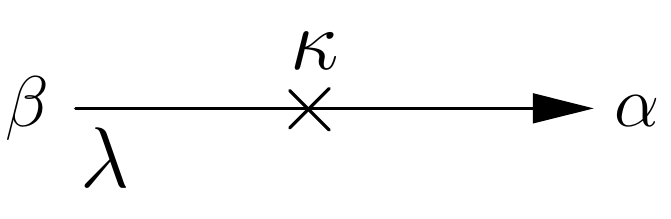}}} \epp
\end{equation}
Then the $U(1)$ symmetry (\ref{u1sym}) becomes
\begin{multline}
     \text{\raisebox{-49pt}{\includegraphics[width=.24\textwidth]{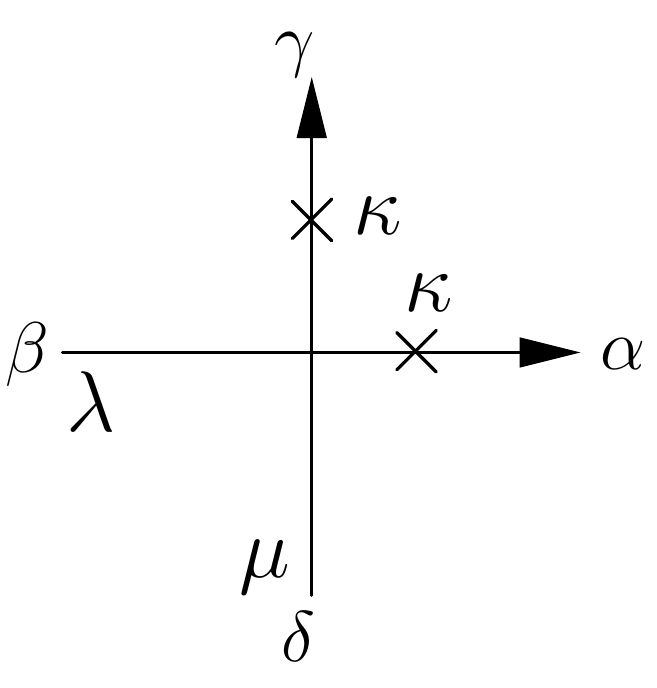}}}
        \: = \: \Th^\a_{\a'} (\k) \Th^\g_{\g'} (\k) R^{\a' \g'}_{\be \de} (\la, \m) \\[-14ex]
        \: = \: R^{\a \g}_{\be' \de'} (\la, \m)
	\Th^{\be'}_{\be} (\k) \Th^{\de'}_{\de} (\k) \: = \:
        \text{\raisebox{-50pt}{\includegraphics[width=.24\textwidth]{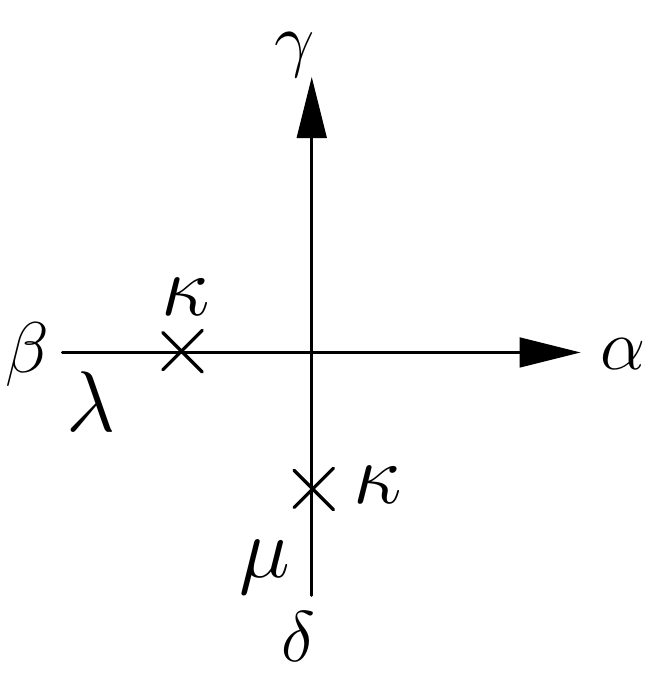}}} \epp
\end{multline}
\item
Examples:
\begin{align}
     & {R^{t_1}}^{\g \a}_{\de \be} (\n, \la) \: = \:
       R^{\de \a}_{\g \be} (\n, \la) \: = \:
       \text{\raisebox{-58pt}{\includegraphics[width=.26\textwidth]{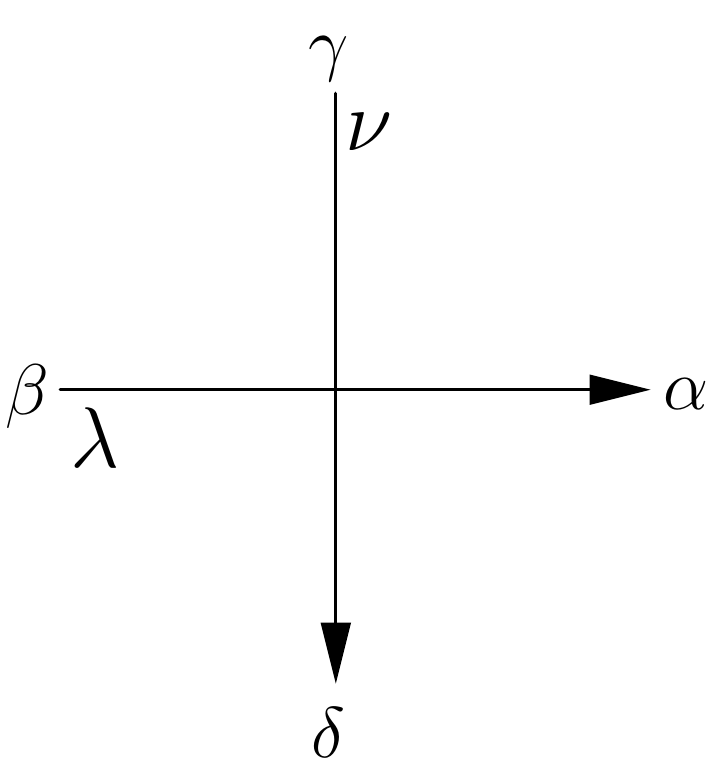}}} \\[-5ex]
     & \Th^\a_{\a'} (\k) {R^{t_1}}^{\g_2 \a'}_{\de_2 \a''} (\n_2, \la)
       R^{\a'' \g_1}_{\be \de_1} (\la, \n_1) \: = \:
       \text{\raisebox{-49pt}{\includegraphics[width=.48\textwidth]{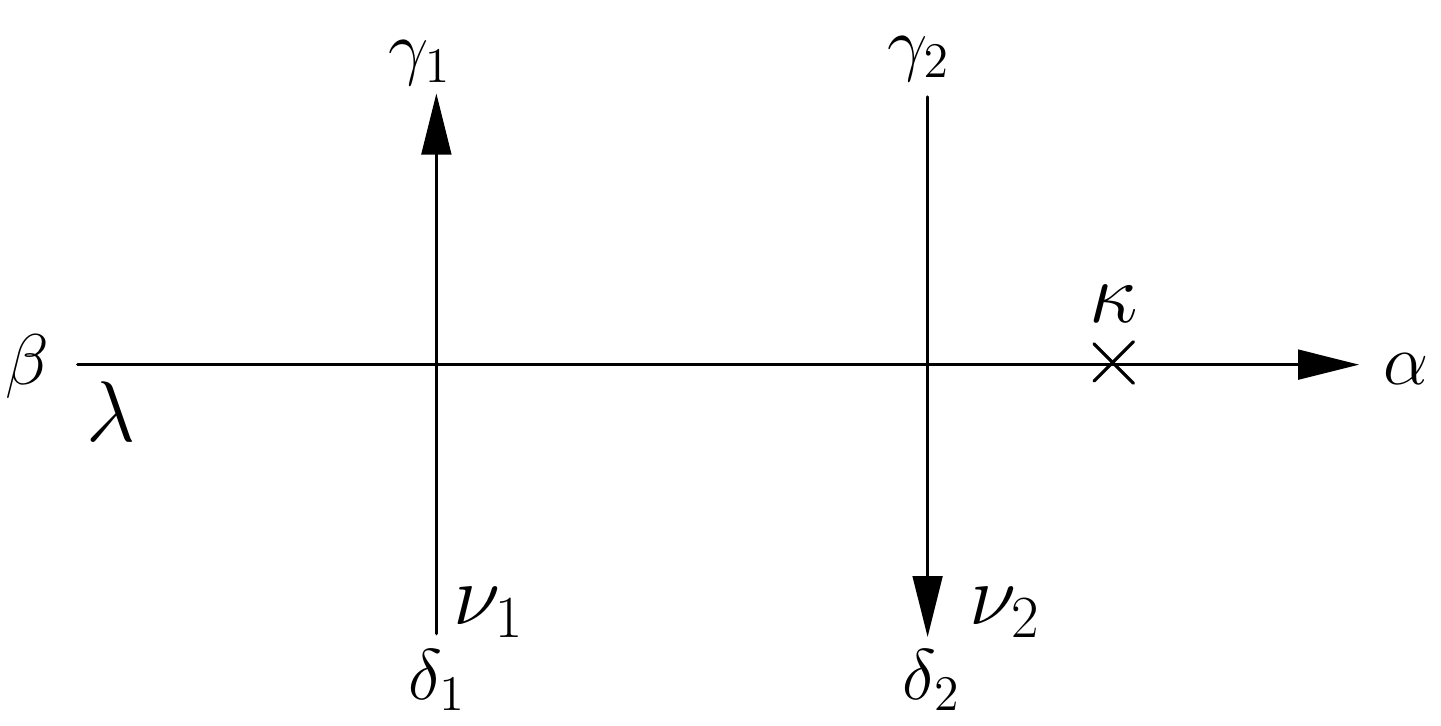}}} \\[1ex]
     & \bigl( \overline{t}_\perp (\la) \bigr)^{\g_{- L + 1} \dots \g_L}_{\de_{- L + 1} \dots \de_L}
       = R^{\g_{- L + 1} \a_{- L + 1}}_{\de_{- L + 1} \a_{- L + 2}} (0, \la)
         R^{\g_{- L + 2} \a_{- L + 2}}_{\de_{- L + 2} \a_{- L + 3}} (0, \la) \dots
         R^{\g_L \a_L}_{\de_L \a_{- L + 1}} (0, \la) \notag \\ & \mspace{130.mu} = \:
         \text{\raisebox{-40pt}{\includegraphics[width=.48\textwidth]{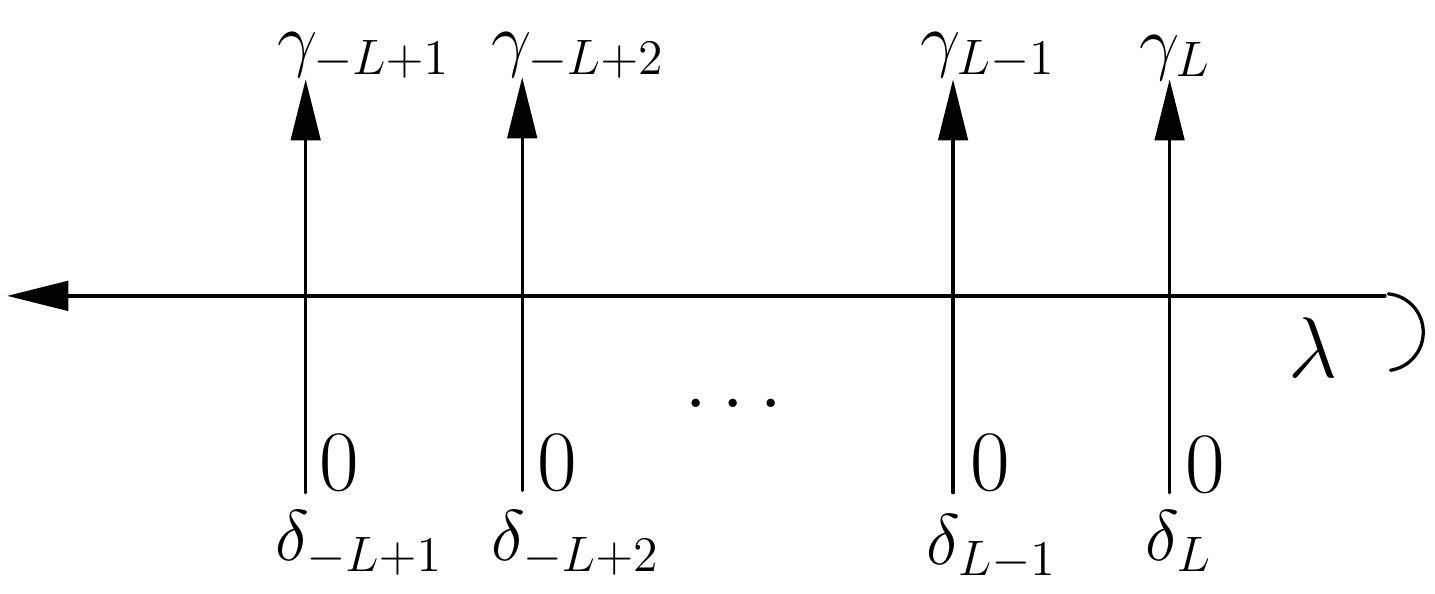}}} \\[1ex]
     & \bigl( t_\perp (\la) \overline{t}_\perp (\m)
              \bigr)^{\g_{- L + 1} \dots \g_L}_{\de_{- L + 1} \dots \de_L} \: = \:
         \text{\raisebox{-55pt}{\includegraphics[width=.48\textwidth]{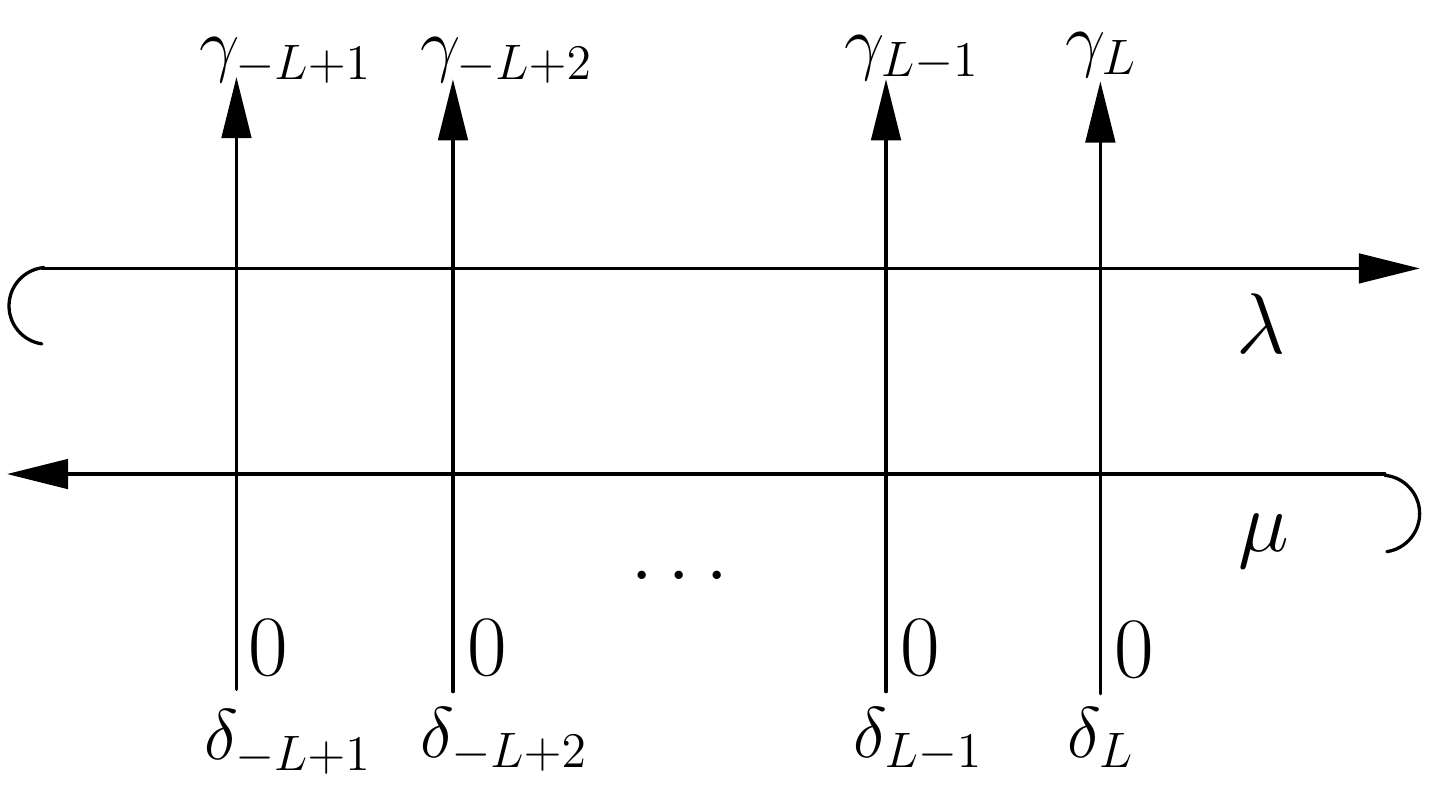}}} \epp
\end{align}
\end{enumerate}
\subsection{Comments}
The graphical notation is generally useful in tensor calculus, not
only in integrable systems. Other examples are Feynman diagrams,
tensor networks and matrix product states.
\section{Partition function, density matrix and static correlation
functions}
\subsection{Statistical operator of the (grand) canonical ensemble}
The graphical notation is appropriate for discussing the various types
of (reduced) density matrices considered in the literature. We start with
the most fundamental one which is the statistical operator of the grand
canonical ensemble. Define
\begin{equation}
     \widetilde \rho_{N, L} = \re^{\k \hat \PH/T}
        \Bigl[ t_\perp \Bigl(- \frac{h_R}{NT}\Bigr)
	       \overline{t}_\perp \Bigl(\frac{h_R}{NT}\Bigr)\Bigr]^\frac{N}{2} \epc \qd
     Z_{N, L} = \tr_{-L+1, \dots, L} \widetilde \rho_{N, L} \epp
\end{equation}
Since, by (\ref{rhoqtm}),
\begin{equation}
     \lim_{N \rightarrow \infty} \widetilde \rho_{N, L} = \re^{- H_L/T} \epc
\end{equation}
we call
\begin{equation}
     \rho_{N, L} = \frac{\widetilde \rho_{N, L}}{Z_{N, L}}
\end{equation}
the finite Trotter number approximant to the statistical operator.

$\widetilde \rho_{N, L}$ has the graphical representation
\begin{multline}
     {\text{$\widetilde \rho$}_{N, L}}^{\g_{- L + 1} \dots \g_L}_{\de_{- L + 1} \dots \de_L}
        \: = \:
         \text{\raisebox{-92pt}{\includegraphics[width=.60\textwidth]{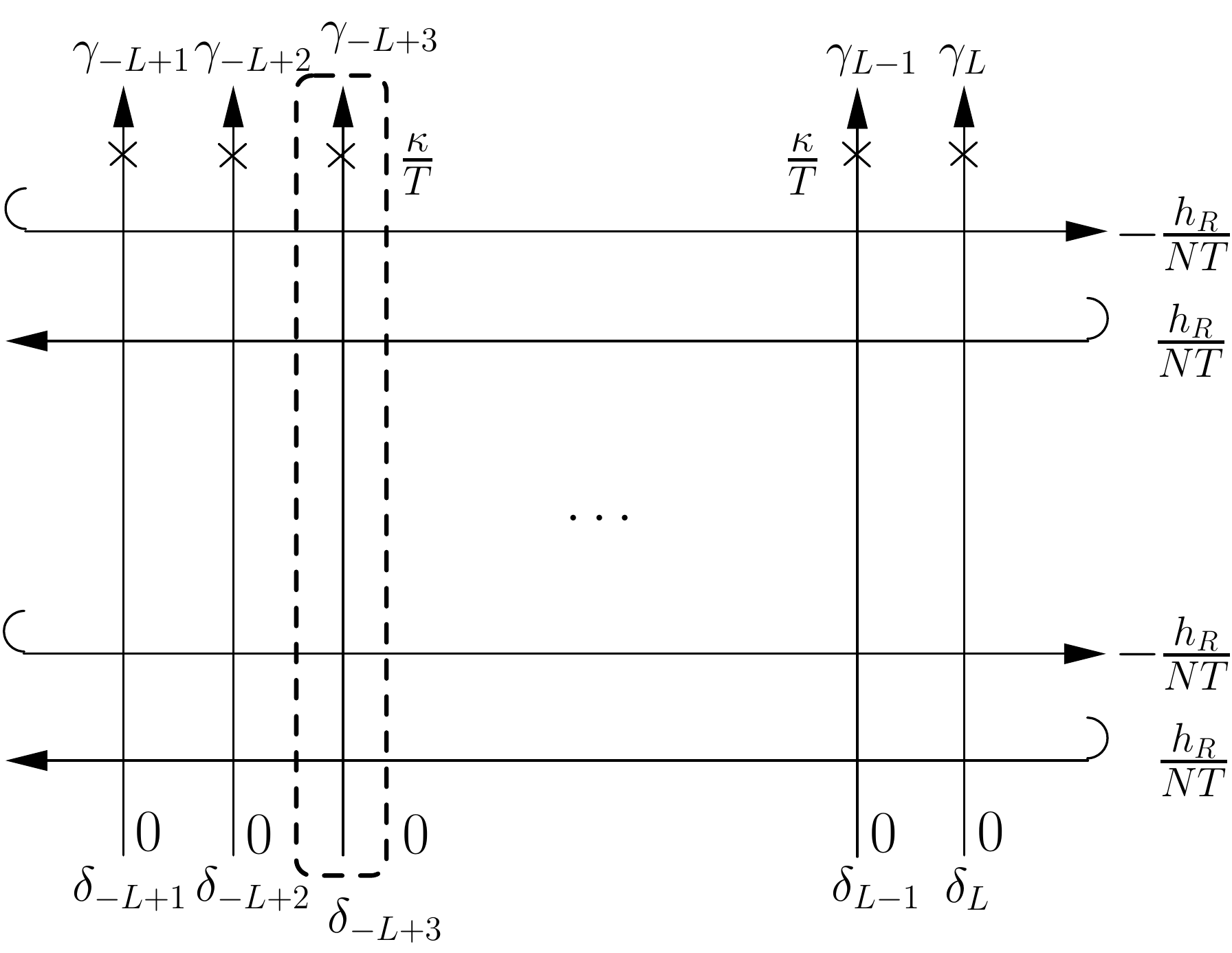}}} \\[1ex]
	= \tr_{\bar 1, \dots, \overline N} \bigr\{ T_{- L + 1} (0) \dots T_L (0) \bigr\} \epp
\end{multline}
Regarding the graph directly and in a reference frame rotated by $\p/2$,
the equality of left and right hand side (an hence the proof of
(\ref{rhoqtm})) becomes obvious and needs no further explanation.
Note that the object in the dashed frame is the staggered, inhomogeneous
monodromy matrix introduced above.
\subsection{Partition function and free energy per lattice site}
The thermodynamics of a quantum spin system is determined by its
partition function
\begin{equation}
     Z_L = \tr_{-L+1, \dots, L} \bigl\{ \re^{- H_L/T} \bigr\}
         = \lim_{N \rightarrow \infty} Z_{N, L}
	 = \lim_{N \rightarrow \infty} \tr_{\bar 1, \dots, \overline N}
	   \bigl\{ \bigl(t(0|\k/T)\bigr)^{2L} \bigr\}
\end{equation}
or by its free energy
\begin{equation}
     F_L = - T \ln Z_L \epc
\end{equation}
respectively.

Let us denote the eigenvalues of $t(\la|\k/T)$ by $\La_n (\la|\k)$,
$n = 0, \dots, d^N - 1$, and assume that they are ordered in such
a way that
\begin{equation} \label{sortlambdas}
     |\La_0 (0|\k)| \ge |\La_1 (0|\k)| \ge |\La_2 (0|\k)| \ge \dots \epp
\end{equation}
If
\begin{enumerate}
\item
the limits $N \rightarrow \infty$ and $L \rightarrow \infty$ commute
and
\item
$|\La_0 (0|\k)/\La_n (0|\k)| < 1$ for all $N \in 2 {\mathbb Z}_+$
and all $n = 1, \dots, d^N - 1$,
\end{enumerate}
the expression of the free energy per lattice site greatly simplifies
in the thermodynamic limit:
\begin{multline} \label{ftl}
     f(T,\k) = \lim_{L \rightarrow \infty} \frac{F_L}{2L}
        = - T \lim_{L \rightarrow \infty} \lim_{N \rightarrow \infty}
	  \frac{1}{2L} \ln \biggl\{ \sum_{n=0}^{d^N-1} \La_n^{2L} (0|\k)\biggr\} \\
        = - T\lim_{N \rightarrow \infty} \ln \La_0 (0|\k) \epp
\end{multline}
It is determined by a sequence of non-degenerate eigenvalues $\La_0$
of the quantum transfer matrix. We shall call these eigenvalues of
largest modulus for fixed $N$ the dominant eigenvalues, the corresponding
eigenvectors $|\k\>$ the dominant eigenvectors.

Proving statements such as the commutativity of the Trotter limit
and the thermodynamic limit usually falls into the realm of
mathematical analysis. Physicists often take a pragmatic attitude
toward more sophisticated mathematical questions, assuming everything
to be alright until a counterexample appears. This is perhaps
why statements (i) and (ii) above have not yet been rigorously
justified. Still we have good reasons to believe that they are
true for all fundamental integrable models and for all $T > 0$.
\begin{enumerate}
\item
Equation (\ref{ftl}) was used to study the thermodynamics of
many fundamental integrable lattice models, most notably of the
Heisenberg XXX and XXZ spin chains \cite{Kluemper93,Suzuki99}
and of the Hubbard model \cite{JKS98c}. These studies have been
compared with other independent methods and give the same results
within the available numerical accuracy. They reproduce the
correct high and low-temperature behaviour and have the correct
free Fermion limits in those cases where they exist.
\item
For the special case of the XXZ chain a rigorous proof
was recently provided for high enough finite temperatures
\cite{GGKS18pp}. Since the proof uses basically only the
locality of the interaction of the model, it is expected
to be generalizable at least to all fundamental integrable
models.
\item
The commutativity of limits was proved for a slightly differently
defined quantum transfer matrix (which is unfortunately less
compatible with the integrable structure imposed by the
Yang-Baxter equation) by M. Suzuki in 1985 \cite{Suzuki85}.
\end{enumerate}
We would like to point out that the existence of the first
limit in (\ref{ftl}), defining the free energy per lattice
site, was proved long time ago (see e.g.\ \cite{Ruelle69}).
Thus, the commutativity of the limits would also imply
the existence of the limit on the right hand side of the
equation.

\begin{remark}
So far we did not use the integrability in any essential way.
What we used was
\begin{enumerate}
\item
the $U(1)$ symmetry
\begin{equation}
     [H, \PH] = 0
\end{equation}
and the regularity
\begin{equation}
     R(\la,\m) = P + (\la - \m) P H^{(2)} + \dots
\end{equation}
where $H^{(2)}$ is a two-site Hamiltonian and the dots denote
terms quadratic in $\la$ and~$\m$.
\end{enumerate}
Simply defining for a given two-site Hamiltonian $H^{(2)}$
\begin{equation}
     R_{H^{(2)}} (\la,\m) = P + (\la - \m) P H^{(2)}
\end{equation}
and using the $U(1)$ symmetry of this object if it exists we
obtain a quantum transfer matrix with dominant eigenvalue
$\La_0 (0|\k)$ and (\ref{ftl}) remains valid. Equation
(\ref{ftl}) can then be used as the starting point for
the implementation of a numerical algorithm for the
calculation of thermodynamic properties of infinite
spin chain systems \cite{Sirker_2002}. The integrability enters the
game only when we calculate $\La_0$.
\end{remark}
\subsection{Density matrix of a chain segment (reduced density
matrix) and expectation values of local operators}
For any $A \in \End {\cal H}_{2L}$ other than the identity
operator there exist $k, l \in \{- L + 1, \dots, L\}$, $k \le l$
and $X \in \End \bigl({\mathbb C}^d\bigr)^{\otimes (l - k + 1)}$
such that $A$ acts non-trivially on sites $k$ and $l$ and
\begin{equation}
     A = X_{k, k+1, \dots, l} \epp
\end{equation}
$X$ is the non-trivial part of $A$. We identify the chain segment
associated with lattice sites $k, \dots, l$ (a certain number of
factors in $\bigl({\mathbb C}^d\bigr)^{\otimes 2L}$) with the
`interval' $[k,l] = (k, k+1, \dots, l)$ and write
\begin{equation}
     X_{k, k+1, \dots, l} = X_{[k,l]} \epp
\end{equation}
The number of sites in $[k,l]$ will be called the length of $X$,
$\ell (X)$. Here $\ell (X) = l - k + 1$. These notions still make
sense for $L \rightarrow \infty$. An operator, whose length stays
finite for $L \rightarrow \infty$ will be called local. For any
local operator of length $m$, with non-trivial part $X$ on the
infinite chain its grand canonical expectation value is
\begin{align} \label{expvallocop}
     \< X_{[k,l]} \> & = \lim_{L \rightarrow \infty}
        \tr_{-L+1, \dots, L} \bigl\{ \r_L (T) X_{[k,l]} \bigr\}
        = \lim_{L \rightarrow \infty}
        \tr_{-L+1, \dots, L} \bigl\{ \r_L (T) X_{[1,\ell(X)]} \bigr\} \notag \\
        & = \lim_{L \rightarrow \infty} \lim_{N \rightarrow \infty}
        \tr_{-L+1, \dots, L} \bigl\{ \r_{N, L} X_{[1,m]} \bigr\} \notag \\
        & = \lim_{L \rightarrow \infty} \lim_{N \rightarrow \infty}
	    \frac{1}{Z_{N, L}} \times
         \text{\raisebox{-99pt}{\includegraphics[width=.55\textwidth]{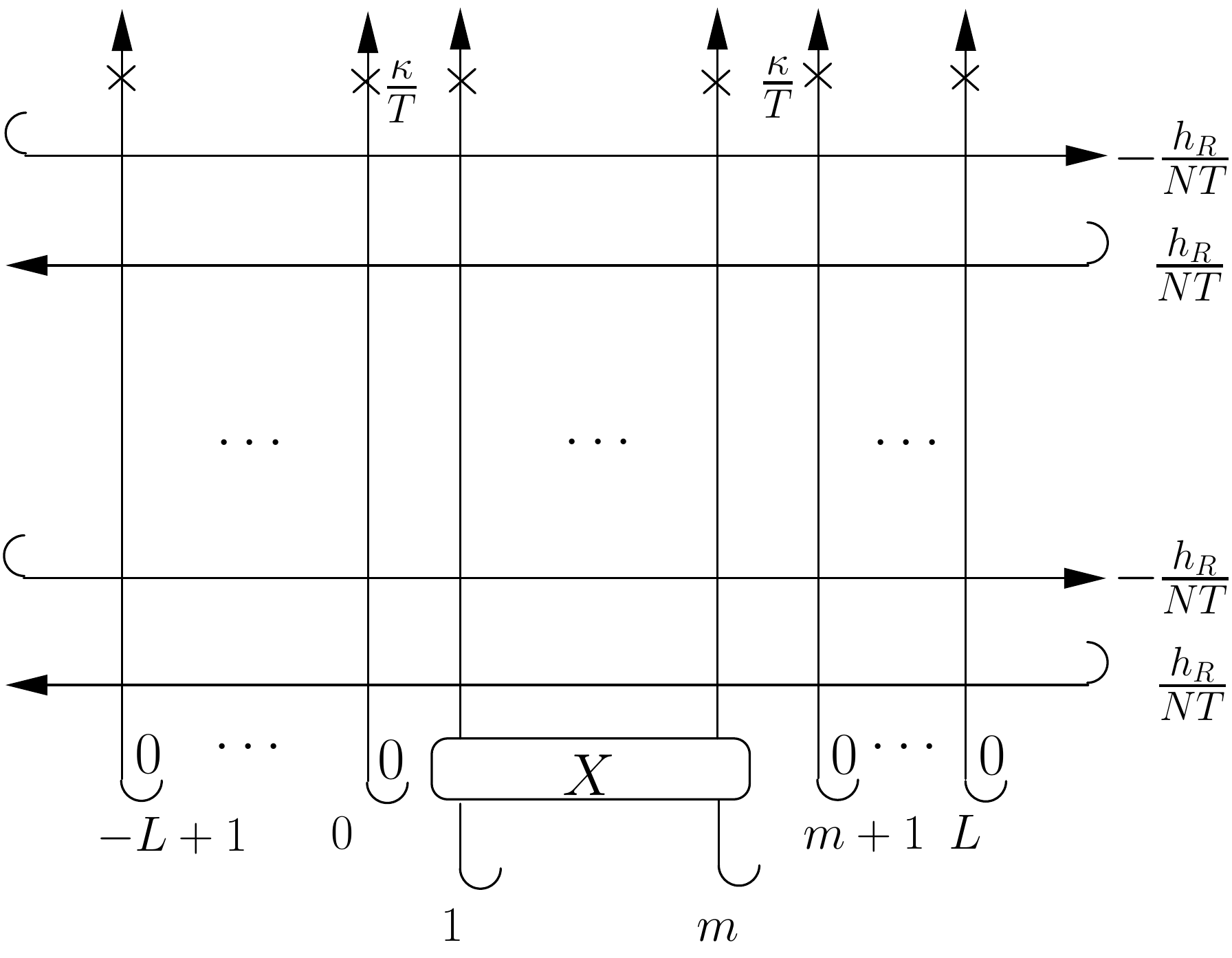}}}
	 \notag \\ & = \lim_{N \rightarrow \infty} \:
	 \text{\raisebox{-92pt}{\includegraphics[width=.52\textwidth]{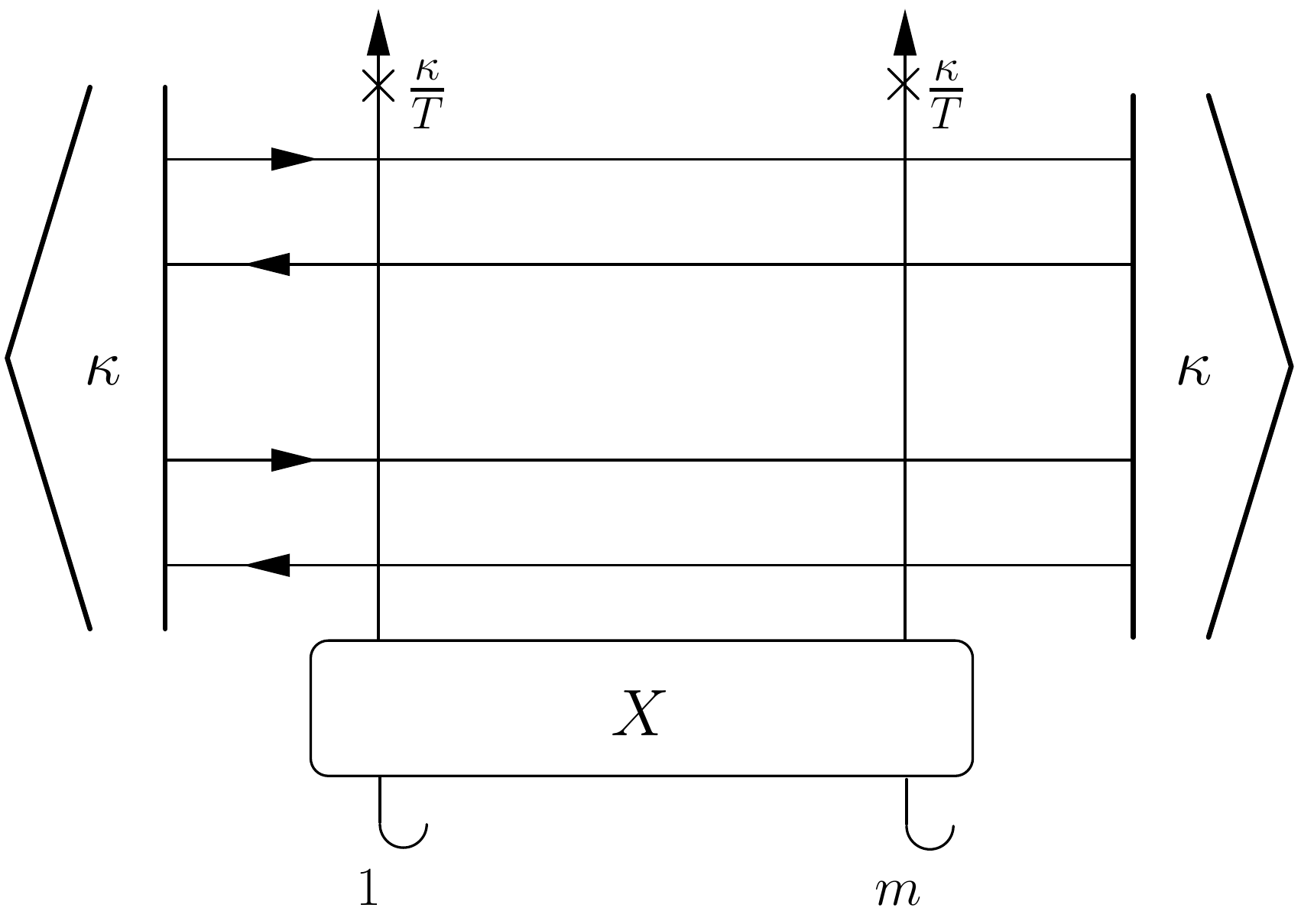}}}
	 \times \frac{1}{\<\k|\k\> \La_0^m (0|\k)} \epp
\end{align}
Here we have used the translation invariance of the Hamiltonian
in the second equation, have inserted the finite Trotter number
approximant to the statistical operator in the third equation,
have represented the resulting expression graphically in the
fourth equation, and have used the fact that the quantum transfer
matrix projects on its dominant state, if applied many times,
in the fifth equation. The latter property holds under the
assumption that points (i) and (ii) below equation (\ref{sortlambdas})
are satisfied. The box around the letter `$X$' represents the
corresponding operator. For the graphical representation of the
dominant state and its dual we use the self-explanatory symbols
$|\k\>$ and $\<\k|$. We see that the expectation values of all local
operators of length $m$ (on the infinite chain) can be calculated
by means of the finite Trotter number approximant to the reduced
density matrix associated with the interval (or chain chain segment)
$[1,m]$,
\begin{equation} \label{defdensmatapp}
     D_m^{(N)} (T, \k) \: = \:
        \text{\raisebox{-68pt}{\includegraphics[width=.52\textwidth]{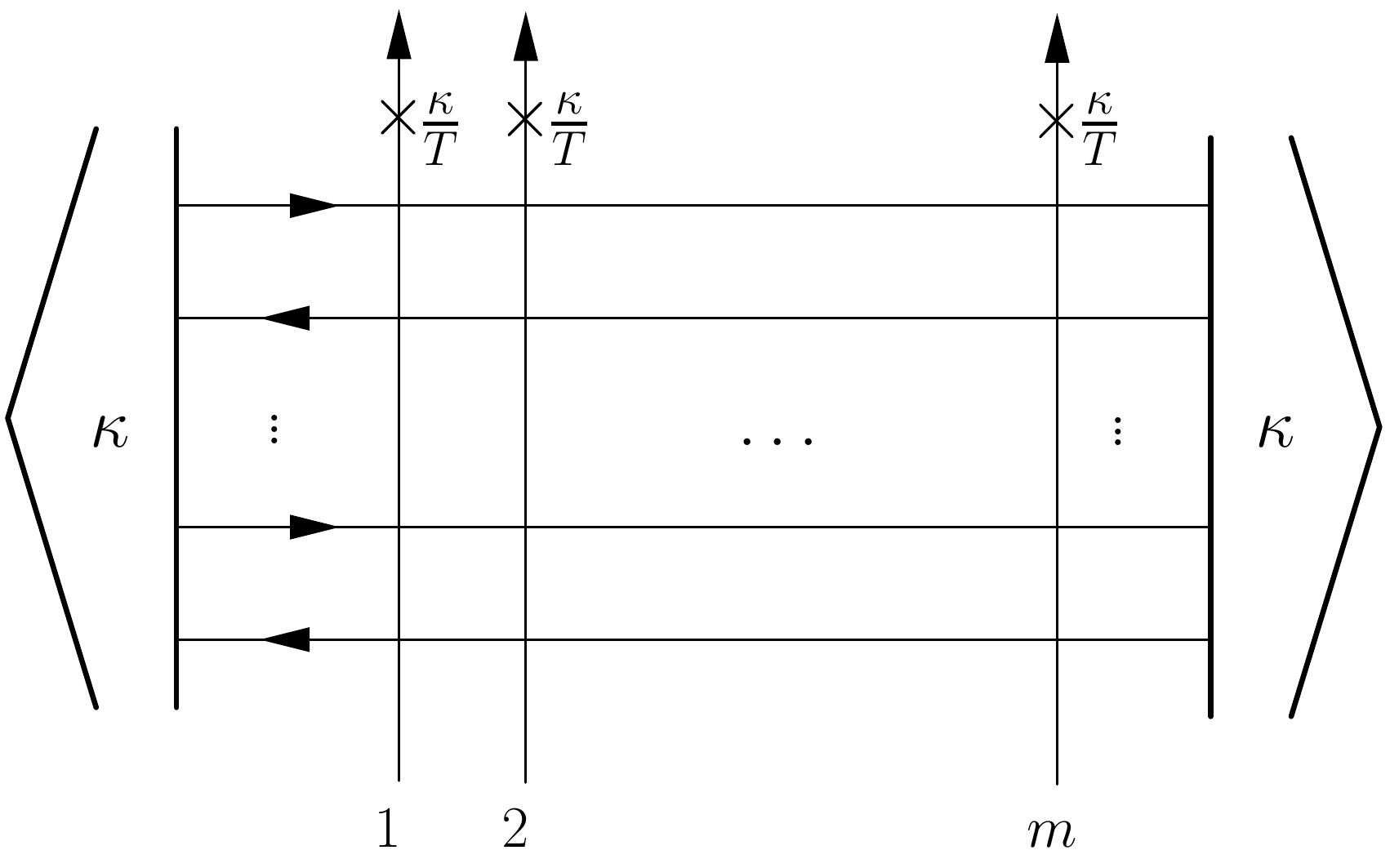}}}
	\times \frac{1}{\<\k|\k\> \La_0^m (0|\k)} \epp
\end{equation}
The reduced density matrix $D_m (T, \k)$ is obtained in the Trotter limit
$N \rightarrow \infty$, and, as can be read off from (\ref{expvallocop}),
\begin{equation} \label{xexp}
     \< X_{[k,l]} \> = \tr_{1, \dots, m} \bigl\{ D_m (T, \k) X \bigr\} \epp
\end{equation}
Using the sequence $D_m (T, \k)_{m \in {\mathbb N}}$ of reduced density
matrices we can calculate the expectation value of any local operator
on the infinite chain (a procedure which is called an `inductive limit').
Equation (\ref{defdensmatapp}) is central for the theory of correlation
functions of integrable models, in particular of the XXZ chain to be
considered in more detail below.
\subsection{Comments}
Coming back to algebraic expressions we see that
\begin{equation} \label{dmalg}
     D_{m+1} (T, \k)
        = \lim_{N \rightarrow \infty}
	  \frac{\<\k|T (0|\k/T)^{\otimes (m+1)}|\k\>}
	       {\<\k|\k\> \La_0 (0|\k)^{(m+1)}} \epp
\end{equation}

For any two ultra-local operators $x, y \in \End {\mathbb C}^d$ we define
$X(\la|\k) = \tr\{x T(0|\k/T)\}$ and $Y(\la|\k) = \tr\{y T(0|\k/T)\}$.
Employing this notation and using (\ref{xexp}) and (\ref{dmalg}) we
obtain the following expression for the two-point correlation function
of $x$ and $y$:
\begin{multline}
     \<x_1 y_{m+1}\> = \lim_{N \rightarrow + \infty}
            \frac{\<\k|\tr\{X(0|\k)\} t (0|\k/T)^{m-1} \tr\{Y(0|\k)\}|\k\>}
            {\<\k|\k\> \La_0 (0|\k)^m} \\[1ex]
       = \lim_{N \rightarrow + \infty} \sum_n
            \frac{\<\k|X(0|\k)|\k, n\>\<\k, n|Y(0|\k)|\k\>}
            {\<\k|\k\> \La_0 (0|\k) \<\k, n|\k, n\> \La_n (0|\k)}
            \biggl(\frac{\La_n (0|\k)}{\La_0 (0|\k)}\biggr)^m \epp
\end{multline}
In the second line we have expanded $t (0|\k/T)^{m-1} \tr\{Y(0|\k)\}|\k\>$
in an eigenbasis $\{|\k, n\>\}$ of $t (0|\k/T)$. The series on the
right hand side is what we call a `thermal form factor series'
\cite{DGK13a}. It is a useful tool for analysing the large-distance
asymptotic behaviour of the two-point function.
\section{The characterisation of reduced density matrices}
\subsection{Expectation values of local operators in pure states}
Using the regularity relation (\ref{reg}) we obtain the following
graphical identities
\begin{equation}
     X_{[1]} \: = \:
        \text{\raisebox{-24pt}{\includegraphics[width=.13\textwidth]{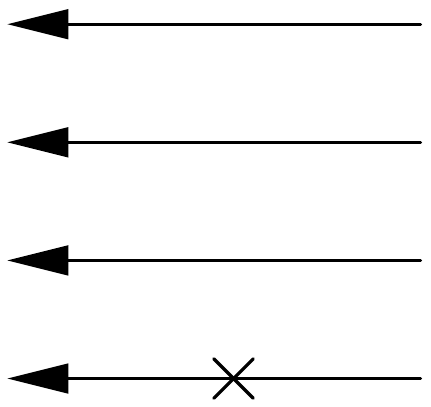}}} \: = \:
        \text{\raisebox{-36pt}{\includegraphics[width=.13\textwidth]{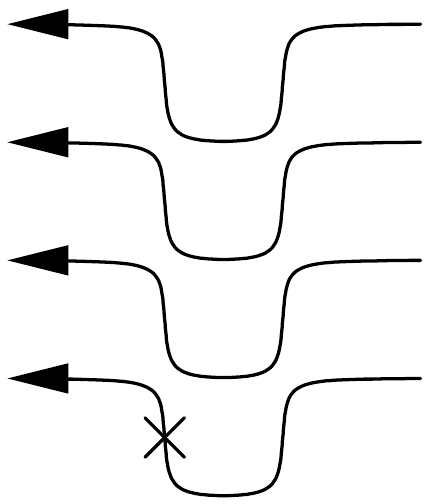}}} \: = \:
        \text{\raisebox{-36pt}{\includegraphics[width=.13\textwidth]{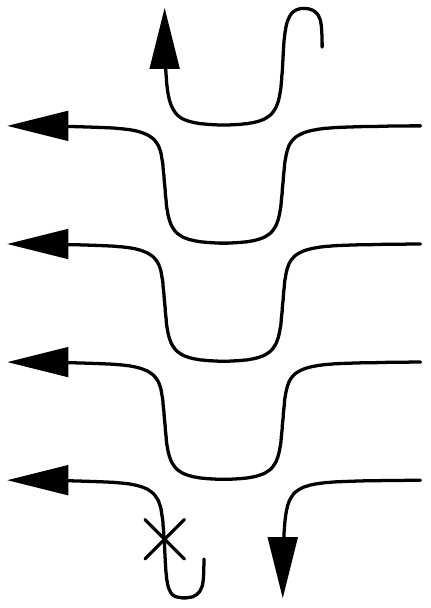}}} \: = \:
        \text{\raisebox{-54pt}{\includegraphics[width=.14\textwidth]{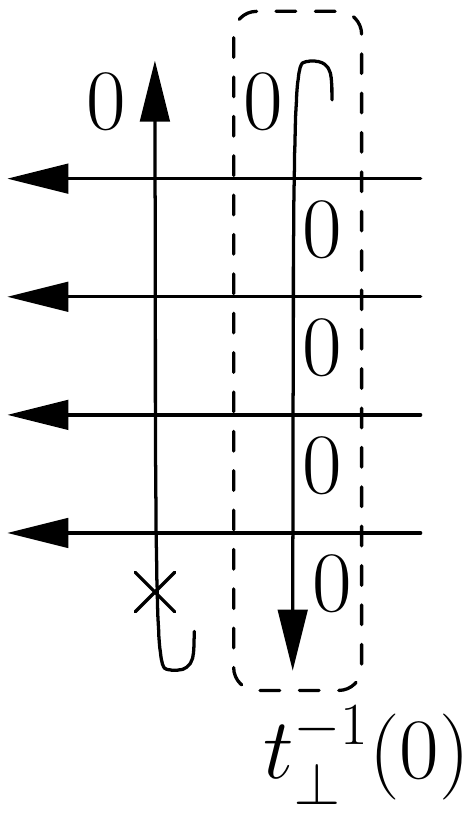}}} \epp
\end{equation}
This generalises to
\begin{align} \label{invprobgraph}
     X_{[1,m]} \: & = \:
        \text{\raisebox{-35pt}{\includegraphics[width=.15\textwidth]{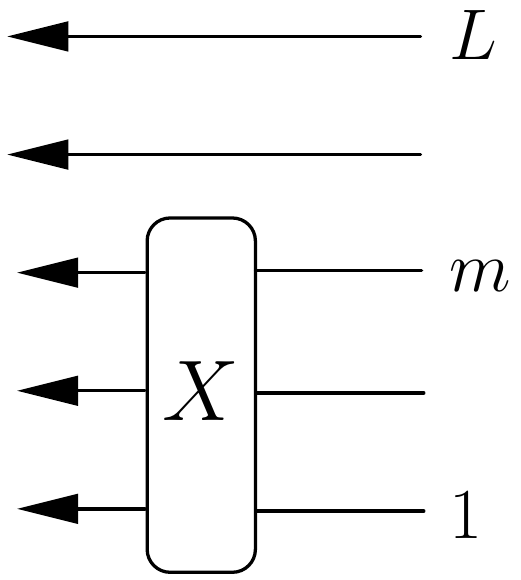}}} \: = \:
        \text{\raisebox{-86pt}{\includegraphics[width=.28\textwidth]{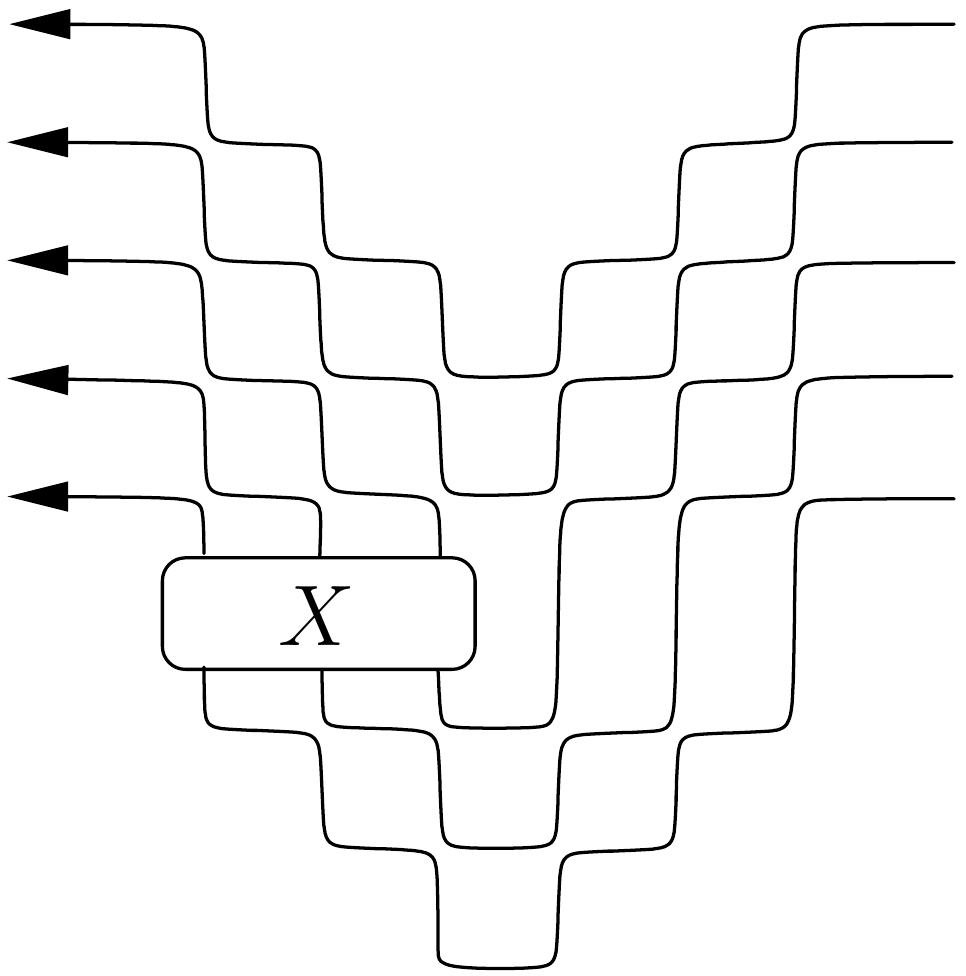}}} \: = \:
        \text{\raisebox{-57pt}{\includegraphics[width=.28\textwidth]{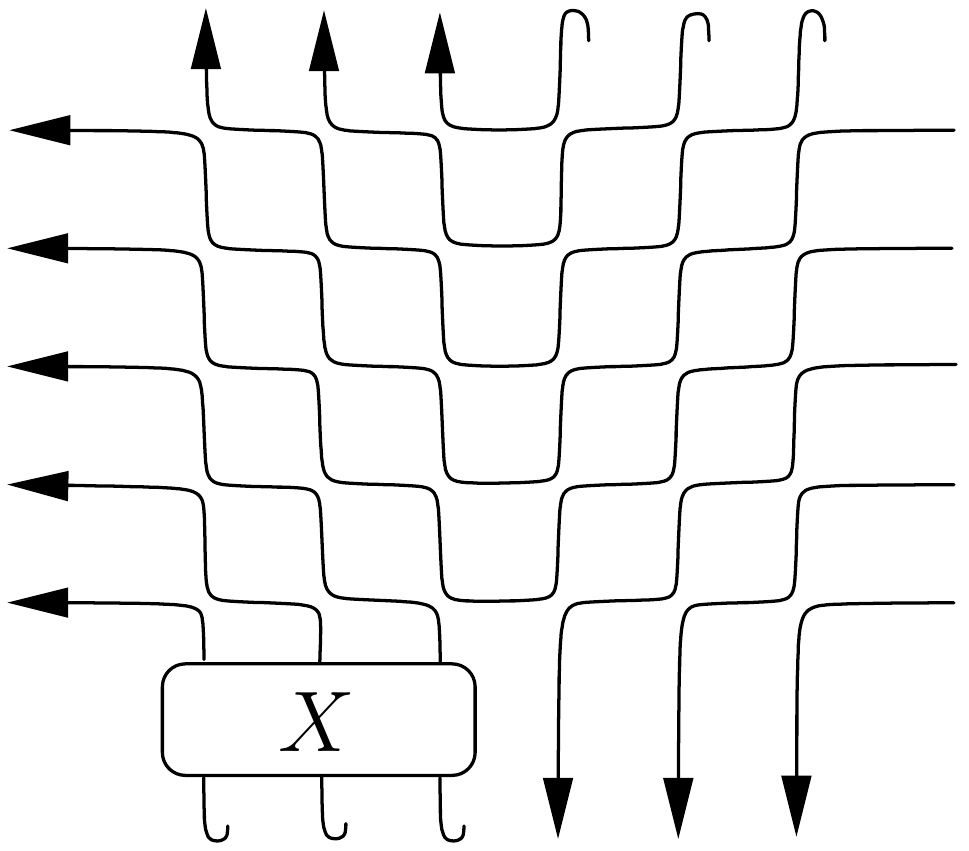}}}
	\notag \\ & = \:
        \text{\raisebox{-67pt}{\includegraphics[width=.32\textwidth]{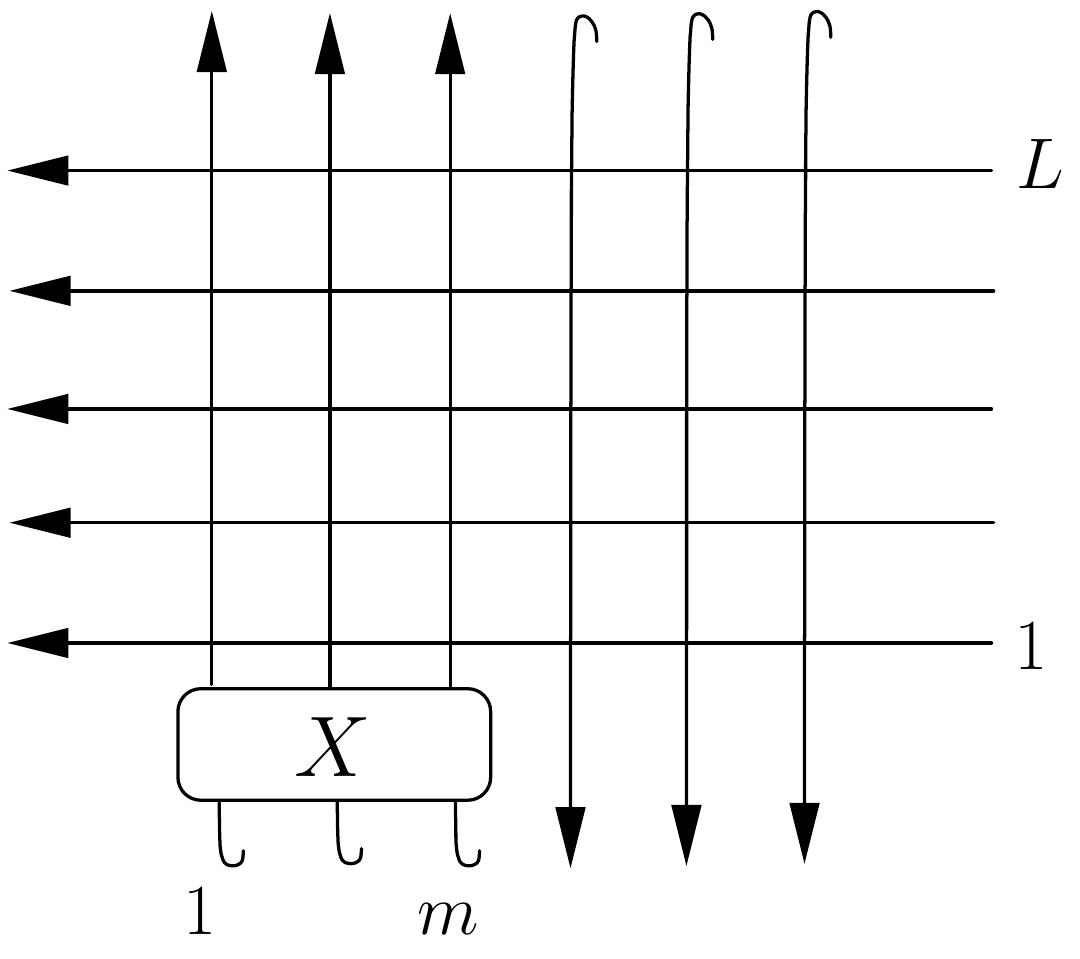}}} \epp
\end{align}
If now $|\Ps\>$ is any eigenstate of $t_\perp (\la)$ with eigenvalue
$\La (\la)$, then (\ref{invprobgraph}) implies that
\begin{equation} \label{densmatpsi}
     \frac{\<\Ps|X_{[1,m]}|\Ps\>}{\<\Ps|\Ps\>} \: = \:
        \text{\raisebox{-36pt}{\includegraphics[width=.19\textwidth]{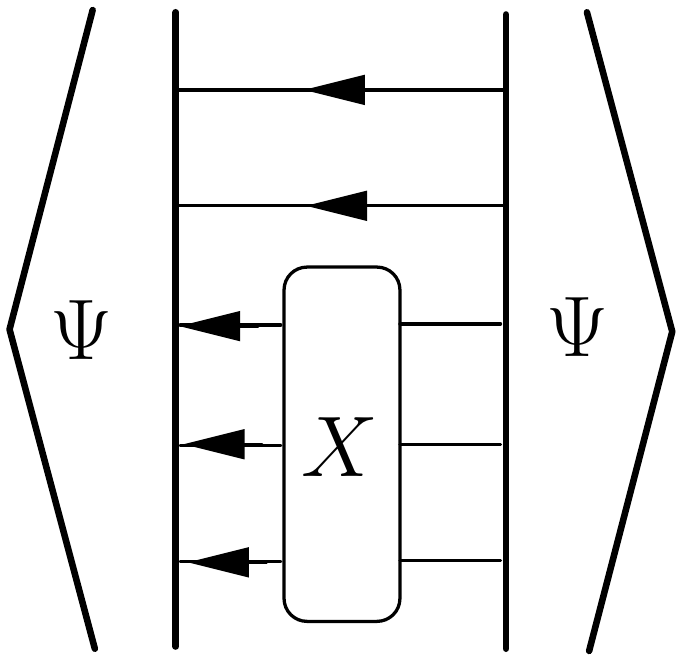}}}
	\: \times \frac{1}{\<\Ps|\Ps\>} \: = \:
        \text{\raisebox{-58pt}{\includegraphics[width=.19\textwidth]{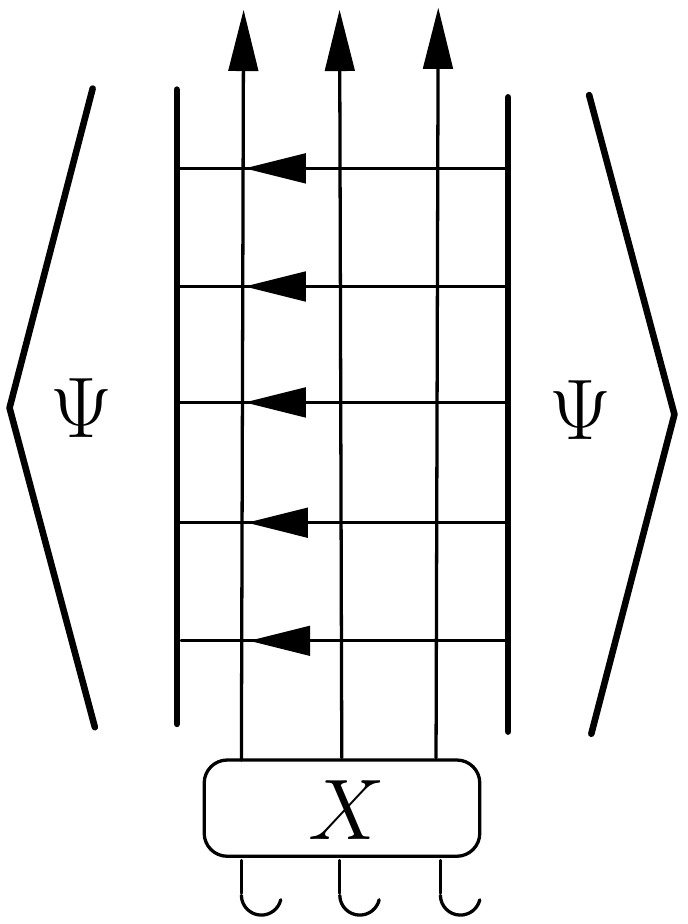}}}
	\: \times \frac{1}{\<\Ps|\Ps\> \La^m (0)} \epp
\end{equation}
These are two graphical representations of the reduced density matrix
of the interval $[1,m]$ associated with the eigenstate $|\Ps\>$.
Remarkably, the expression on the right hand side is of the same
form as in (\ref{expvallocop}) (a twist can also be included).
\begin{remark}
We would like to emphasise the following:
\begin{enumerate}
\item
We still did not use integrability here. The construction works for
non-integrable models as well.
\item
The above is a graphical version of `the solution of the inverse
problem' \cite{KMT99a,GoKo00,MaTe00}.
\item
The problem of calculating the reduced density matrix from
(\ref{defdensmatapp}) or (\ref{densmatpsi}) is still largely unsolved,
even for integrable models. Most of the results available in the
literature refer to a few simple example systems related to the
spin-$\frac 12$ XXZ chain on which we concentrate in the following.
\end{enumerate}
\end{remark}
\subsection{Further generalisations of the reduced density matrix
with the example of the XXZ chain}
An idea in the spirit of Baxter \cite{Babook}, which was rather helpful
for actually calculating the reduced density matrix of a chain segment,
was to generalise the definition by attaching spectral parameters to
the vertical lines \cite{JMMN92,JiMi96} and a twist to one of the
states \cite{BJMST06b}. We shall denote the corresponding un-normalised
reduced density matrix by
\begin{equation} \label{unnorm}
     {\cal D}^{(N)} (\x_1, \dots, \x_m|T, \k, \k') \: = \:
        \text{\raisebox{-67pt}{\includegraphics[width=.52\textwidth]{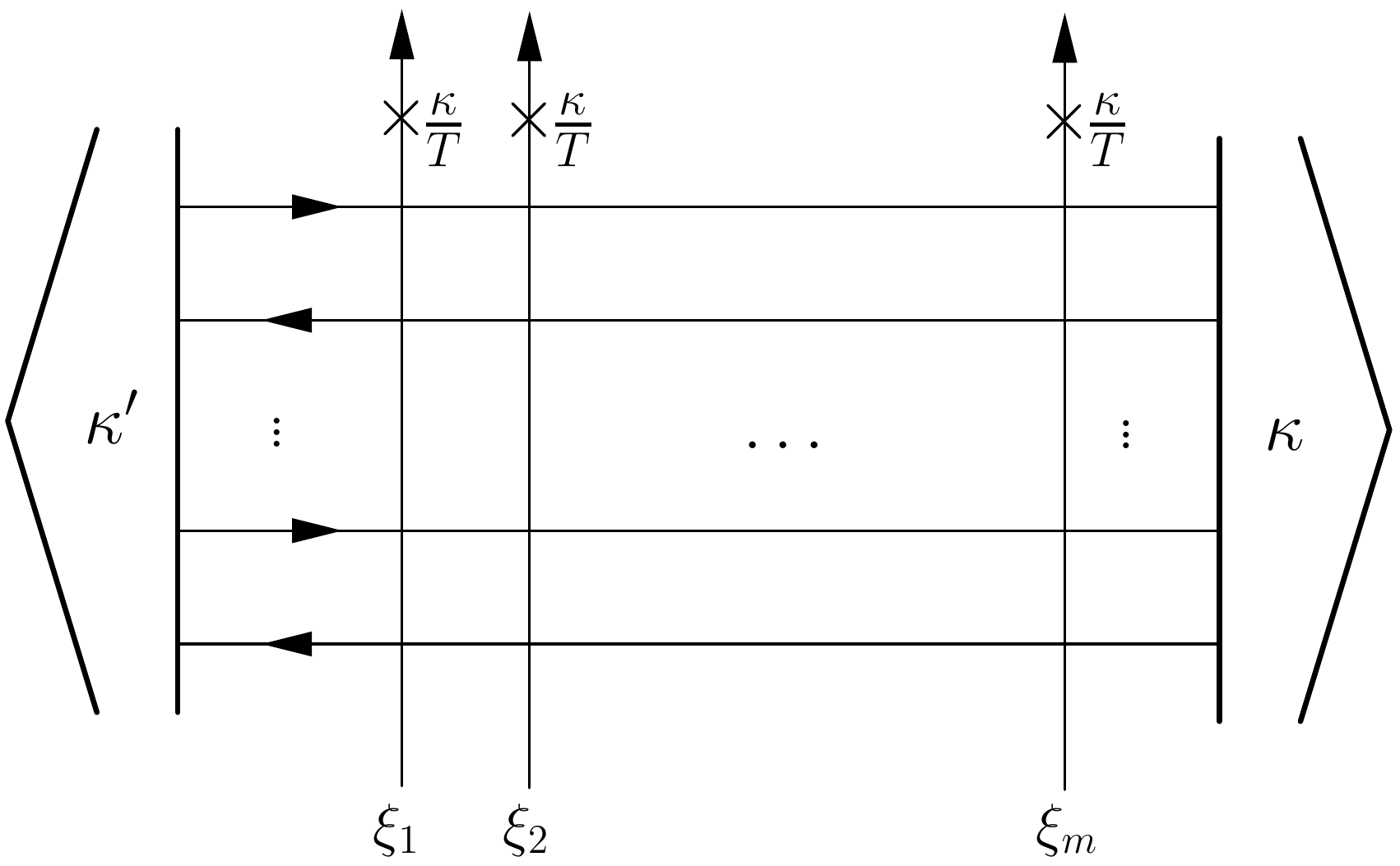}}}
\end{equation}
With this we may define the twisted, inhomogeneous finite Trotter number
approximant to the reduced density matrix,
\begin{equation} \label{normdensmat}
     D^{(N)} (\x_1, \dots, \x_m|T, \k, \k') =
        \frac{{\cal D}^{(N)} (\x_1, \dots, \x_m|T, \k, \k')}
             {\tr_{1, \dots, m}
	      \bigl\{{\cal D}^{(N)} (\x_1, \dots, \x_m|T, \k, \k')\bigr\}} \epp
\end{equation}
It determines the physical reduced density matrix as
\begin{subequations}
\begin{align}
     & D_m (T, \k) = D (0, \dots, 0|T, \k, \k) \epc \\[1ex]
     & D (\x_1, \dots, \x_m|T, \k, \k') = \lim_{N \rightarrow \infty}
       D^{(N)} (\x_1, \dots, \x_m|T, \k, \k') \epp \label{densmatintw}
\end{align}
\end{subequations}

It turns out that the new parameters $\x_j$, $j = 1, \dots, m$, and
$\a = (\k' - \k)/T$ regularise the mathematical expressions for $D^{(N)}$.
The parameter $\a$ acquires a physical meaning in certain scaling limits,
e.g.\ in the conformal limit \cite{BJMS10}.

For the XXZ spin-$\frac 12$ chain $D^{(N)} (\x_1, \dots, \x_m|T, \k, \k')$
has been calculated and characterised in several different ways (for
finite $N$ and in the Trotter limit).
\begin{enumerate}
\item
By an $m$-fold integral \cite{BoGo09}.
\item
By `discrete rqKZ equations' \cite{AuKl12}.
\item
By an `exponential form', involving a double integral and the
annihilation part of the so-called Fermionic basis \cite{BJMST08a}.%
\footnote{This so far only works in the restricted situation, when
the magnetic field $h = 0$. The general case would require the
completion of the algebra by a bosonic annihilation operator
postulated in \cite{BoGo09}.}
\item
By the `JMS theorem' using the creation part of the Fermionic
basis \cite{JMS08}.
\end{enumerate}
Every single of these charactersations is technically involved
and would justify a series of lectures on its own. For this reason
we can only provide a brief description, which will necessarily
stay somewhat vague, and a short guide to the literature.

The first explicit description of the reduced density matrix
of a chain segment of length $m$ of the XXZ chain was obtained
in \cite{JMMN92} and had form of an $m$-fold multiple integral.
It was derived for the ground state of the spin chain in the
massive antiferromagnetic regime. The derivation relied on
the construction of representations of a deformed vertex-operator
algebra.

Subsequently the multiple-integral formula was rederived and
generalised by different methods and different authors. An
extension to the massless groundstate phase at vanishing
magnetic field was obtained in \cite{JiMi96}. The derivation
relied on the use of functional equations of qKZ-type
\cite{FrRe92,Smirnov92b} that had been introduced before
in order to characterise form factors of integrable massive
quantum field theories.

A derivation by Bethe Ansatz \cite{KMT99b} made it possible
to take into account a finite magnetic field and opened
the way to treat the finite temperature and finite length
cases in \cite{GKS05,DGHK07}. A multiple-integral representation
for the most general inhomogeneous and twisted case
(\ref{densmatintw}) was eventually obtained in \cite{BoGo09}.
The non-vanishing density matrix elements are of the form
\begin{multline} \label{multint}
     D^{\e_1' \dots \e_m'}_{\e_1 \dots \e_m}
           (\x_1, \dots, \x_m|T, \k, \k') =
	   \biggl[ \prod_{j=1}^p \int_{\cal C} \rd m(\la_j) \:
	          F^+_{\ell_j} (\la_j) \biggr]
	   \biggl[ \prod_{j=p+1}^m \int_{\cal C} \rd \overline{m}(\la_j) \:
	          F^-_{\ell_j} (\la_j) \biggr] \\[1ex]
           \frac{\det_{j, k = 1, \dots, m} \bigl[- G(\la_j, \x_k|\k, \k') \bigr]}
	        {\prod_{1 \le j < k \le m} \sh(\la_j - \la_k - \h)
		 \sh(\x_k - \x_j)} \epc
\end{multline}
where we have used the notation
\begin{align} \label{defmeasureetc}
     \rd m(\la) &
        = \frac{\rd \la}{2 \p \i \, \r(\la|\k, \k') (1 + \fa (\la, \k))} \epc \qd
     \rd \overline{m} (\la)
        = \fa (\la, \k) \rd m(\la) \epc \\[2ex] \notag
     F_{\ell_j}^\pm (\la) &
        = \prod_{k=1}^{\ell_j - 1} \sh(\la - \x_k) \mspace{-9.mu}
	  \prod_{k=\ell_j + 1}^m \mspace{-9.mu} \sh(\la - \x_k \mp \h) \epc \qd
     \ell_j = \begin{cases}
                 \e_j^+ & j = 1, \dots, p \\
		 \e_{m - j + 1}^- & j = p + 1, \dots, m
               \end{cases}
\end{align}
with $\e_j^+$ the $j$th plus in the sequence $(\e_j)_{j=1}^m$, $\e_j^-$
the $j$th minus sign in the sequence $(\e_j')_{j=1}^m$ and $p$ the
number of plus signs in $(\e_j)_{j=1}^m$.

The definition of the integration contour ${\cal C}$ depends
on which parameter regime is considered. For $|\D|< 1$, for
instance, we may choose a rectangle centered around the origin
of the complex plane with sides parallel to the real and imaginary
axes and of height $|\h| - 0_+$ and large finite width.
The functions $\fa$ and $G$ are solutions of convolution type
integral equations with respect to the contour ${\cal C}$ and
integration kernel
\begin{equation}
     K_\a (\la) = \re^{- \a} \cth (\la - \h) - \re^\a \cth (\la + \h) \epp
\end{equation}
The function $\fa$ solves the non-linear integral equation
\begin{equation} \label{nlietemhom}
     \ln (\fa (\la, \k)) = - \frac{2\k}{T} - \frac{2J \sh^2 (\h)}{T \sh(\la) \sh(\la + \h)}
        - \int_{\cal C} \frac{\rd \m}{2 \p \i}
		        K_0 (\la - \m) \ln (1 + \fa (\m, \k )) \epc
\end{equation}
whereas $G$ is the solution of the linear integral equation
\begin{multline} \label{newg}
     G(\la, \x|\k ,\k') = \re^{-\a} \cth(\la - \x - \h) - \r (\x|\k, \k') \cth (\la - \x) \\
                  + \int_{\cal C} \rd m(\m) K_\a (\la - \m) G(\m, \x)|\k ,\k') \epp
\end{multline}
Here and in (\ref{defmeasureetc})
\begin{equation} \label{defrho}
     \r (\x|\k, \k') = \frac{\La_0 (\x|\k')}{\La_0 (\x|\k)}
\end{equation}
is the ratio of dominant eigenvalues with different twist parameters.

The multiple-integral formula (\ref{multint}) is compact and
memorizable but, as any true multiple integral, not very
efficient for the actual computation of the density matrix elements
\cite{BoGo05}. Remarkably, however, it turned out that the multiple
integrals factorise into sums over products of single integrals.
This was explictly worked out with examples \cite{BoKo01,BoKo02,%
BGKS06,BoGo09}. It motivated efforts to directly calculate the
density matrix elements in factorised form. This is what is the
main point behind items (ii)-(iv) above. In \cite{BJMST04a,BJMST04b}
a reduced form of the qKZ equation (rqKZ) was derived for the ground
state of the XXX and XXZ chains and solved in a form corresponding
to the factorised integrals. In \cite{AuKl12} the equation was
generalised to the finite temperature case. The effort to uncover
the structure behind the factorisation led to the discovery of
the Fermionic basis in \cite{BJMST08a,BJMS09a} and eventually to
a proof of the factorisation under very general conditions
(`JMS theorem' \cite{JMS08}). The latter claims that all density
matrix elements and hence all static correlation functions can
be expressed in terms of only two basic functions, the function
$\r$, equation (\ref{defrho}), and a function $\om$ which for
the inhomogeneous finite temperature case was characterised in
terms of the functions $\fa$ and $G$, equations (\ref{nlietemhom}), 
(\ref{newg}), in \cite{BoGo09}:
\begin{multline}
     \re^{- \a (\x_1 - \x_2)} \om (\x_1, \x_2|\k, \k') =
        2 \Ps (\x_1, \x_2|\k, \k') + K_\a (\x_1 - \x_2) \\
	+ \bigl(\r(\x_1|\k, \k') - \r(\x_2|\k, \k')\bigr) \cth(\x_1 - \x_2) \epc
\end{multline}
where
\begin{multline}
     \Ps (\x_1, \x_2) =
        \int_{\cal C} \rd m(\la) \: G(\la, \x_2|\k, \k') \\ \times
	\bigl(\re^\a \cth(\la - \x_1 - \h) - \r(\x_1|\k, \k') \cth(\la - \x_1) \bigr) \epp
\end{multline}

Factorisation has been used to calculate short-range static
correlation functions of the XXZ and XXX chains
(see e.g.\ \cite{SABGKTT11,MiSm19} and references listed
therein). Here is an example for the XXX chain at
$T = 0$, $h = 0$ that was first obtained in \cite{BoKo02},
\begin{multline}
     {D_4}^{1111}_{1111} (0,0) = \\ \frac{1}{5} - 2 \ln 2 + \frac{173}{60} \z (3)
        - \frac{11}{6} \z(3) \ln 2 - \frac{51}{80} \z^2 (3) - \frac{55}{24} \z (5)
	+ \frac{85}{24} \z (5) \ln 2 \epp
\end{multline}
The Riemann $\z$-functions arise from the function $\om$ in
the necessary limits. A more generic example of a finite
temperature correlation function of the XXZ chain is
\begin{equation}
     \< \s_1^x \s_2^x \> = - \frac{\om (0,0|\k, \k)}{2 \sh(\h)}
        + \frac{\ch(\h) \6_\la \6_\a \om (\la, 0|\k + \a T, \k)\bigr|_{\la = 0, \a = 0}}{2}
	\epp
\end{equation}
Factorisation becomes rapidly inefficient for operators $X$
of length $\ell (X) \simeq 10$ or larger as the number of terms in
the factorised form of the correlation functions grows very rapidly.
In particular, it was so far inefficient for the calculation of
the asymptotics of multi-point correlation functions. Yet, for
these other methods, based e.g.\ on effective field theoretical
descriptions \cite{LuTe03} or on the use of form factor series
\cite{KKMST11b}, are available.

It is an interesting open question, if factorisation of static
correlation functions is a peculiarity of the XXZ quantum
spin chain or rather a generic feature of integrable lattice
models.
\subsection{Properties of the generalised density matrix}
While the derivation of the multiple-integral formula and the
construction of the Fermionic basis are rather technical and
certainly exceed the scope of these lecture notes, a characterisation
of the reduced density matrix by its properties is comparatively
easy.
\begin{enumerate}
\item
The `normalisation condition',
\begin{equation}
     \tr_{1, \dots, m} \bigl\{D^{(N)} (\x_1, \dots, \x_m|T, \k, \k')\bigr\} = 1 \epc
\end{equation}
is obvious from the definition (\ref{normdensmat}).
\item
The `exchange relation',
\begin{multline}
     \check R_{j-1, j} (\x_{j-1}, \x_j) D^{(N)} (\x_1, \dots, \x_m|T, \k, \k') \\
        = D^{(N)} (\x_1, \dots, \x_{j-2}, \x_j, \x_{j-1}, \x_{j+1}, \dots, \x_m|T, \k, \k')
	  \check R_{j-1, j} (\x_{j-1}, \x_j) \epc
\end{multline}
where $\check R(\la, \m) = P R(\la, \m)$, is a consequence of the
Yang-Baxter equation and can most easily be seen using its
graphical form (\ref{graphybe}).
\item
The `left and right reduction properties',
\begin{subequations}
\begin{align} \label{rightred}
     & \tr_m \bigl\{D^{(N)} (\x_1, \dots, \x_m|T, \k, \k')\bigr\} =
        D^{(N)} (\x_1, \dots, \x_{m-1}|T, \k, \k')\bigr\} \epc \\
     & \tr_1 \bigl\{\Th_1 (\a) D^{(N)} (\x_1, \dots, \x_m|T, \k, \k')\bigr\} =
       \notag \\ & \mspace{216.mu}
        \r(\x_1|\k, \k') D^{(N)} (\x_2, \dots, \x_m|T, \k, \k') \epc
	\label{leftred}
\end{align}
\end{subequations}
are again obvious from (\ref{unnorm}) and (\ref{normdensmat}).
\end{enumerate}
These properties hold for any fundamental model (with $U(1)$-symmetry).
Further properties follow from special properties of the
$R$-matrix, as e.g.\ group invariance, asymptotics as a function of
the spectral parameter, crossing symmetry.

\begin{remark}
The left reduction fixes the one-point functions connected with
the $U(1)$-symmetry. Setting $m=1$ and taking the derivative of
(\ref{leftred}) with respect to $\k'$ at $\k' = \k$ we obtain
\begin{equation}
     \tr_1 \bigl\{ \hat \ph_1 D^{(N)} (\x_1|T, \k, \k) \bigr\}
        = T \6_\k \ln \bigl( \La_0 (\x_1|\k) \bigr) \epp
\end{equation}
Sending $N \rightarrow \infty$, setting $\x_1 = 0$ and using
(\ref{ftl}) we conclude that
\begin{equation}
     \< \hat \ph_1 \> = - \6_\k f(T, \k)
\end{equation}
which is consistent with the standard thermodynamic relation.
\end{remark}
\subsection{Comments}
Using the graphical notation it is not difficult to explain the
reduced qKZ equation which can be used in order to calculate the
reduced density matrix of the XXZ chain. We define
\begin{equation}
     Y_{1, \dots, n+1} \: = \:
        \text{\raisebox{-68pt}{\includegraphics[width=.47\textwidth]{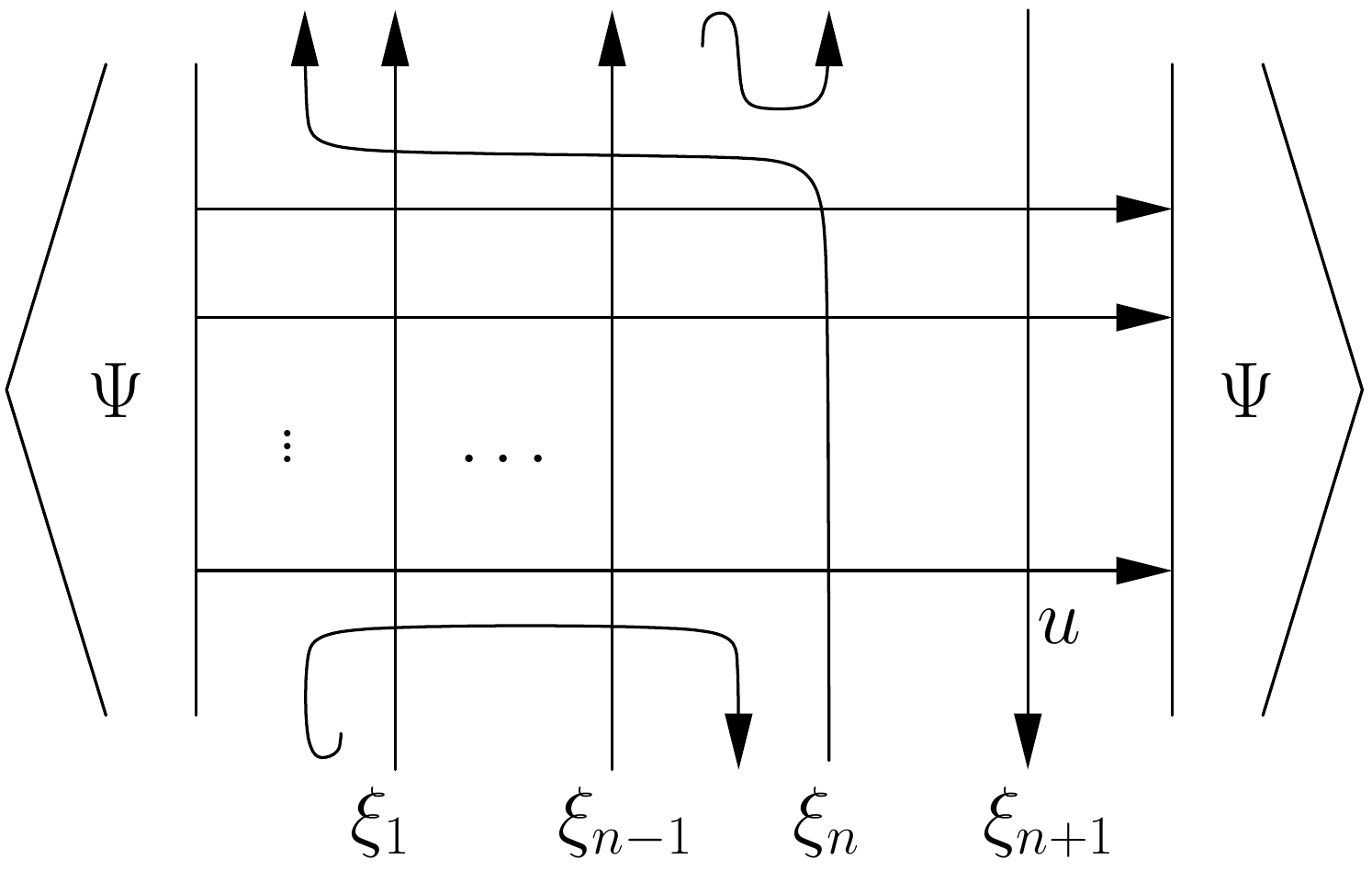}}} \epp
\end{equation}
Here we suppose that there are arbitrarily many horizontal
lines and the spectral parameter on the lowest line is $u$.
We further assume that $|\Ps\>$ is a transfer matrix eigenstate.
Then
\begin{equation}
     \tr_n Y_{1, \dots, n+1}\bigr|_{\x_n = \x_{n+1} = u}
        = \tr_{n+1} Y_{1, \dots, n+1}\bigr|_{\x_n = \x_{n+1} = u}
\end{equation}
by construction. It follows that
\begin{multline}
     \text{\raisebox{-66pt}{\includegraphics[width=.40\textwidth]{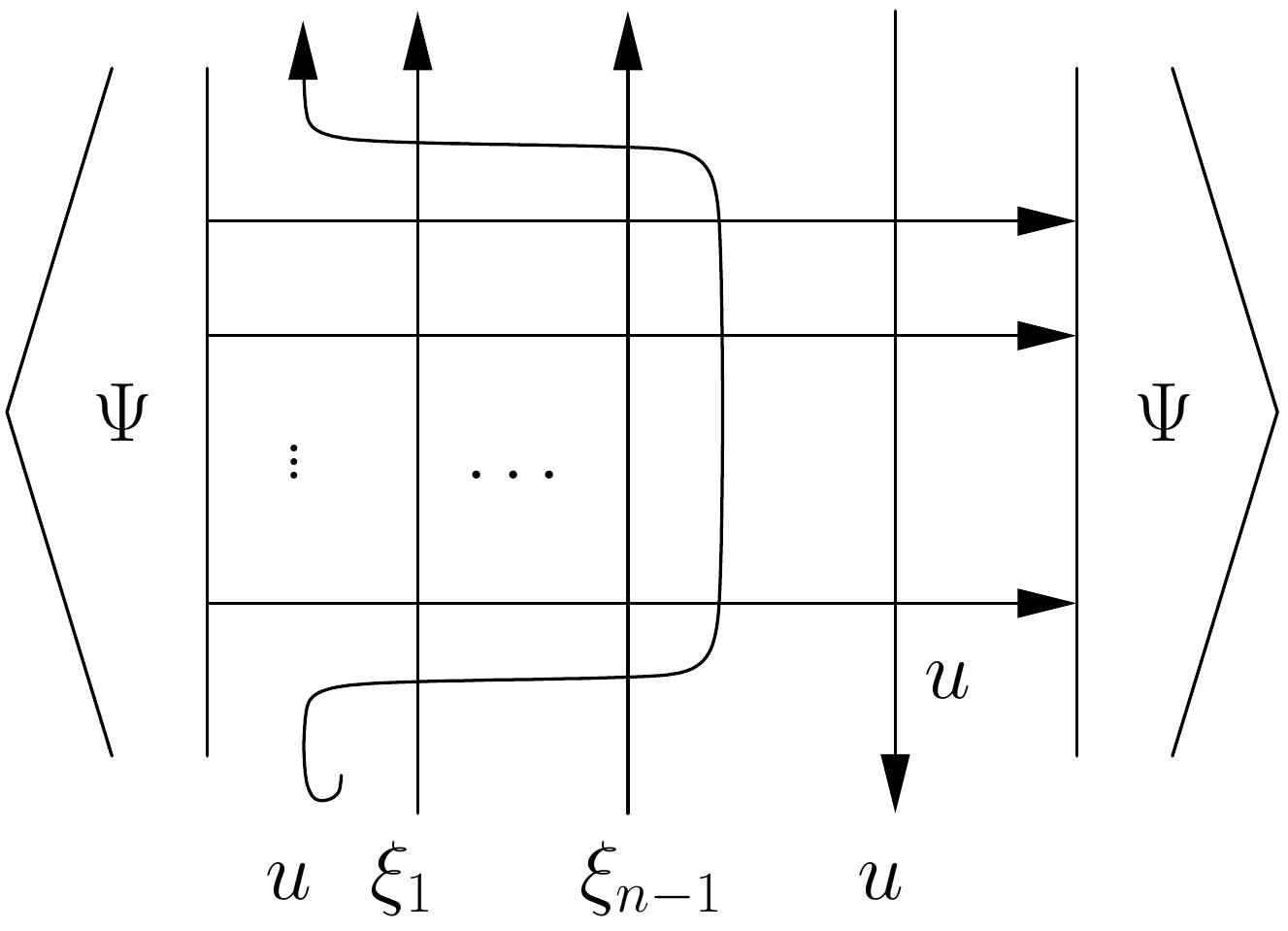}}} \: = \:
     \La (u) \cdot
     \text{\raisebox{-66pt}{\includegraphics[width=.40\textwidth]{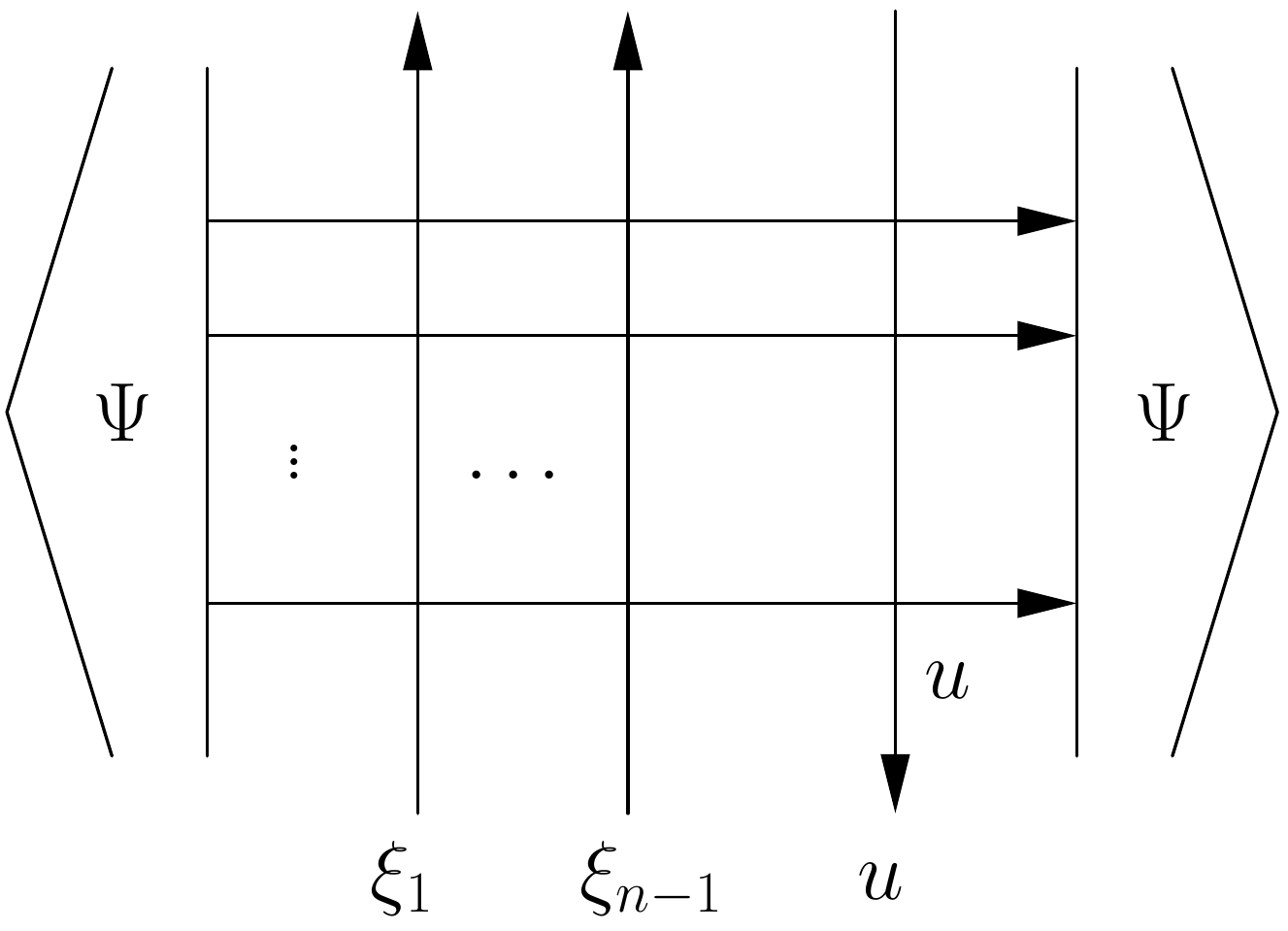}}}
     \\[2ex] = \: \frac{1}{\La (u)} \cdot
     \text{\raisebox{-66pt}{\includegraphics[width=.40\textwidth]{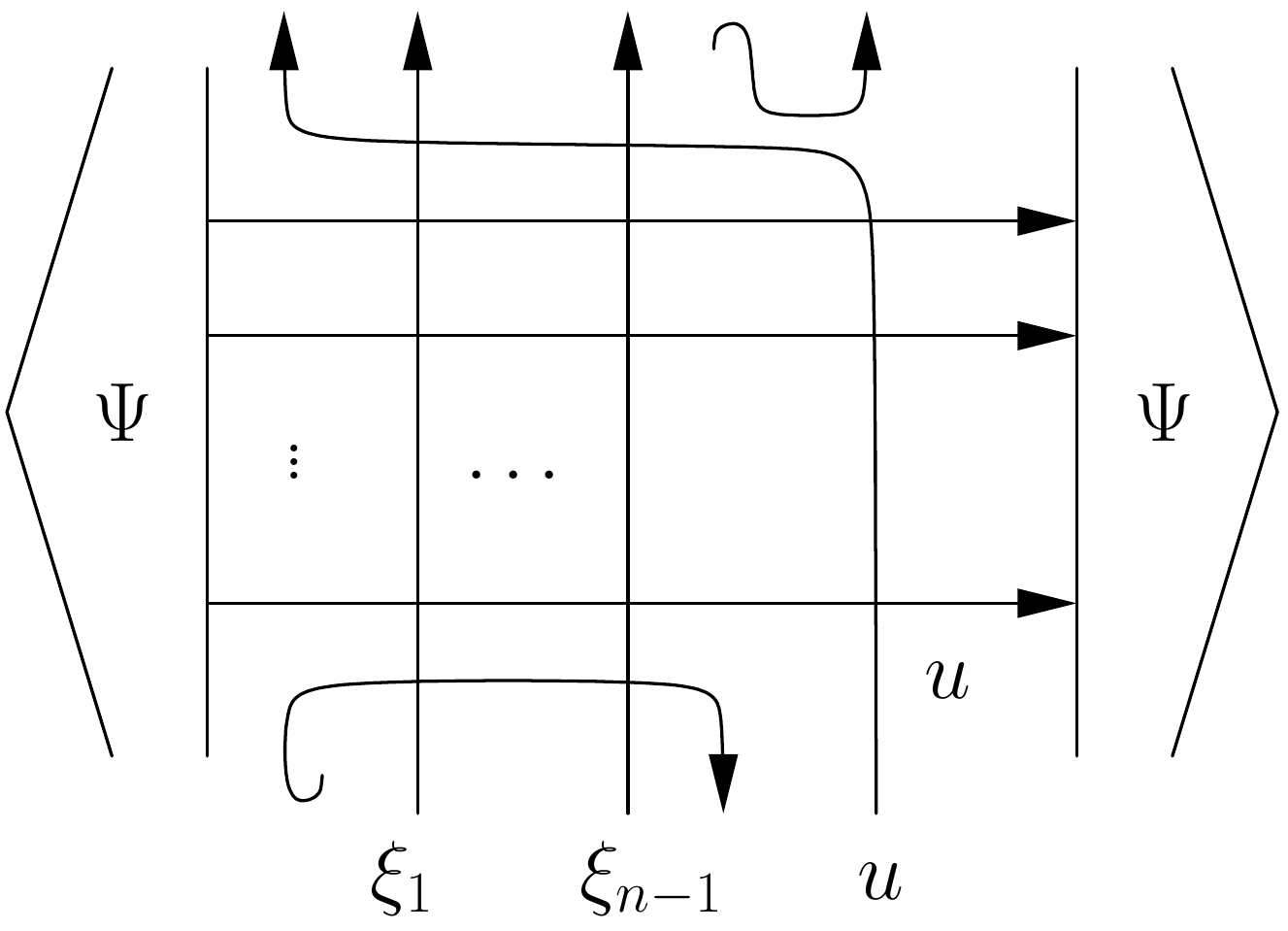}}} \epp
\end{multline}
If the $R$-matrix exhibits crossing symmetry, like in case of the XXZ chain, then
the arrow direction of the rightmost transfer matrix on the right hand side of
the first equation can be reversed and the second equation can be interpreted
as a discrete version \cite{AuKl12} of the reduced qKZ equation \cite{BJMST04a}.

\section{Bethe Ansatz and nonlinear integral equation for the
quantum transfer matrix of the XXZ chain}
\subsection{Algebraic Bethe Ansatz for the quantum transfer
matrix}
In this section we consider the staggered and twisted inhomogeneous
monodromy matrix (\ref{stimonodromy}), where $R$ is the $R$-matrix
(\ref{rxxz}) of the XXZ spin-$\frac 12$ chain. Then the auxiliary
space `$a$' is two-dimensional and $T_a$ can be interpreted as a
$2 \times 2$ matrix with operator-valued entries acting on
${{\mathbb C}^2}^{\otimes N}$,
\begin{equation}
     T_a (\la|\a) = \begin{pmatrix}
                       A(\la) & B (\la) \\ C(\la) & D (\la)
		    \end{pmatrix}_a \epp
\end{equation}
The Yang-Baxter algebra relations (\ref{yba}) are a set of quadratic
relations for these entries. These relations allow one to construct
a set of eigenvectors of the quantum transfer matrix
\begin{equation}
     t(\la|\a) = A(\la) + D(\la)
\end{equation}
generated over a pseudo vacuum $|0\>$ which has the properties
\begin{equation}
     C(\la)|0\> = 0 \epc \qd
     A(\la)|0\> = a(\la)|0\> \epc \qd D(\la)|0\> = d(\la)|0\>
\end{equation}
for some complex functions $a(\la)$, $d(\la)$.

The existence of a pseudo vacuum is a non-trivial requirement.
There are representations of the Yang-Baxter algebra which
do not have a pseudo vacuum. For the quantum transfer matrix
of the XXZ chain, however, a pseudo vacuum does exist. This
can be easily inferred from the structure of the $R$-matrices
composing the staggered monodromy matrix (\ref{stimonodromy}).
They take the form
\begin{subequations}
\begin{align}
     R_{a, j} (\la, \n) & =
        \begin{pmatrix}
	   {e_j}^1_1 + b(\la - \n) {e_j}^2_2 & c(\la - \n) {e_j}^1_2 \\
	   c(\la - \n) {e_j}^2_1 & b(\la - \n) {e_j}^1_1 + {e_j}^2_2
	\end{pmatrix}_a \epc \\[1ex]
     {R^{t_1}}_{j,a} (\n, \la) & =
        \begin{pmatrix}
	   {e_j}^1_1 + b(\n - \la) {e_j}^2_2 & c(\n - \la) {e_j}^2_1 \\
	   c(\n - \la) {e_j}^1_2 & b(\n - \la) {e_j}^1_1 + {e_j}^2_2
	\end{pmatrix}_a \epp
\end{align}
\end{subequations}
Setting
\begin{equation}
     |0\> = \bigl( e_1 \otimes e_2 \bigr)^{\otimes N/2}
\end{equation}
we see that
\begin{equation}
     T_a (\la|\a) |0\> =
        \begin{pmatrix}
	   a(\la) & B (\la) \\ 0 & d (\la)
	\end{pmatrix}_a |0\> \epc
\end{equation}
where
\begin{equation}
     a(\la) = \re^{\a/2} \prod_{j=1}^{N/2} b(\n_{2j} - \la) \epc \qd
     d(\la) = \re^{- \a/2} \prod_{j=1}^{N/2} b(\la - \n_{2j-1}) \epp
\end{equation}
For the density matrix of the grand canonical ensemble we set
as before $\n_{2j-1} = h_R/(NT)$, $\n_{2j} = - h_R/(NT)$, $j = 1, \dots, N/2$,
and $\a = \k/T$.

For any set $\{\la\} = \{\la_j\}_{j=1}^M \subset {\mathbb C}$ we
define
\begin{align} \label{defq}
     & Q(\la|\{\la\}) = \prod_{j=1}^M \sh(\la - \la_j) \epc \\
     & |\{\la\}\> = B(\la_M) \dots B(\la_1) |0\> \epp \label{boffshell}
\end{align}
Then we have the following
\begin{theorem*} Algebraic Bethe Ansatz \cite{STF79}.
\begin{equation}
     t(\la|\a) |\{\la\}\> = \La(\la|\{\la\})|\{\la\}\>
\end{equation}
with
\begin{equation}
     \La(\la|\{\la\}) = \frac{a(\la) Q(\la - \h|\{\la\}) + d(\la) Q(\la + \h|\{\la\})}
                             {Q(\la|\{\la\})}
\end{equation}
if $\{\la\}$ is chosen in such a way that the `Bethe Ansatz equations'
\begin{equation} \label{bae}
     \frac{d(\la_j) Q(\la_j + \h|\{\la\})}{a(\la_j) Q(\la_j - \h|\{\la\})} = - 1 \epc
\end{equation}
$j = 1, \dots, M$, are satisfied.
\end{theorem*}

\begin{remark}
All eigenstates are of Bethe Ansatz form (\ref{boffshell}), (\ref{bae})
and form a basis, if $\a$, $\n_j$, $j = 1, \dots, M$, are generic \cite{TaVa95}.
\end{remark}
\subsection{Auxiliary functions}
We may assume that $\a$ and the $\n_j$ are generic. Otherwise we slightly
change their values to make them generic. Then all eigenstates and
eigenvalues of the quantum transfer matrix $t(\la|\a)$ can be labeled
by solutions $\{\la_j^{(n)}\}_{j=1}^{M_n}$ of the Bethe Ansatz equations
(\ref{bae}). Inserting these solutions back into (\ref{defq}) we define
\begin{equation}
     Q_n (\la) = \prod_{j=1}^{M_n} \sh(\la - \la_j^{(n)})
\end{equation}
and the auxiliary functions
\begin{equation} \label{defaux}
     \fa_n (\la) = \frac{d(\la) Q_n (\la + \h)}{a(\la) Q_n (\la - \h)} \epp
\end{equation}
By construction these functions have the important property that
\begin{equation} \label{subsi}
     1 + \fa_n (\la_j^{(n)}) = 0 \epc \qd j = 1, \dots, M_n \epp
\end{equation}
\subsection{Nonlinear integral equations}
Equation (\ref{subsi}) allows us to characterise the auxiliary functions
by means of nonlinear integral equations. Let ${\cal C}_n$ be a simple
closed contour that encircles $\{\la_j^{(n)}\}_{j=1}^{M_n}$ and the
$N/2$-fold pole of $\fa_n$ at $- \frac{h_R}{NT}$, but no other poles
or roots of $1 + \fa_n$. Then the following `monodromy condition'
holds,
\begin{equation} \label{monocondition}
     \int_{{\cal C}_n} \frac{\rd \la}{2 \p \i} \:
        \6_\la \ln \bigl(1 + \fa_n (\la) \bigr) = M_n - \frac{N}{2} = - s_n \epp
\end{equation}
We shall call $s_n$ the `(pseudo-) spin of the $n$th excited state'.

In order to avoid case differentiations we restrict the parameter
$\h$ from now on to $\h = - \i \g$, $\g \in (0,\p/2]$. We would, like
to point out, however, that a similar analysis is possible in all
physically relevant parameter regimes. Define
\begin{equation}
     \e_0^{(N)} (\la) = \k - \frac{TN}{2} \ln \biggl(
	\frac{\sh(\la - \frac{h_R}{NT})}{\sh(\la + \frac{h_R}{NT})}
	\frac{\sh(\la + \frac{h_R}{NT} + \h)}{\sh(\la - \frac{h_R}{NT} + \h)}\biggr)
\end{equation}
and
\begin{equation}
     K (\la) = \cth(\la - \h) - \cth(\la + \h) \epp
\end{equation}
Assuming that $\la \pm \i \g$ is outside the contour ${\cal C}_n$
defined above, for all $\la \in {\cal C}_n$ and for all $\m$ on
and inside ${\cal C}_n$ we obtain the identity
\begin{figure}
\begin{center}
\includegraphics[width=.58\textwidth]{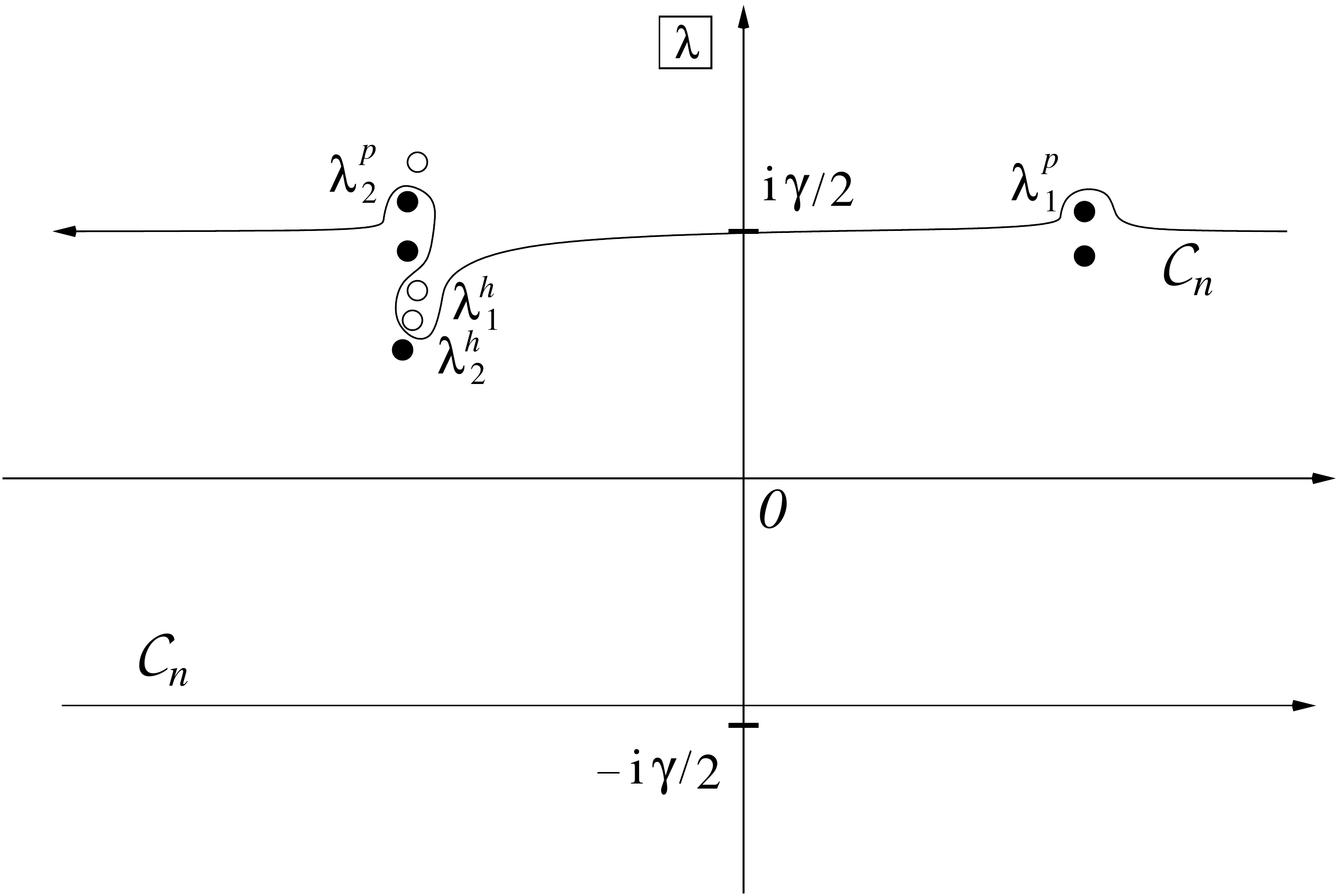}
\caption{\label{fig:cn} Sketch of the contour ${\cal C}_n$. It includes
the point $- \frac{h_R}{NT}$, which is close to the origin for large
enough Trotter number, and all Bethe roots. A few of them, the black
dots, are shown in the figure. More of them are located close to the
origin. The white dots depict roots of $1 + \fa_n$ which are no Bethe
roots (so-called holes). They are all outside ${\cal C}_n$.
}
\end{center}
\end{figure}
\begin{align} \label{derivenlie}
     & \int_{{\cal C}_n} \frac{\rd \m}{2 \p \i} \:
        \ln \biggl(\frac{\sh(\h + \la - \m)}{\sh(\h - \la + \m)}\biggr)
	\6_\m \ln \bigl(1 + \fa_n (\m)\bigr) \notag \\[1ex] & \mspace{36.mu} =
	\sum_{j=1}^{M_n} \ln \biggl(
	\frac{\sh(\h + \la - \la_j^{(n)})}{\sh(\h - \la + \la_j^{(n)})}\biggr)
	- \frac{N}{2} \ln \biggl(
	\frac{\sh(\h + \la + \frac{h_R}{NT})}{\sh(\h - \la - \frac{h_R}{NT})}\biggr)
	\notag \\[1ex] & \mspace{36.mu}
	= \ln \bigl(\fa_n (\la)\bigr) + \frac{\e_0^{(N)} (\la)}{T} - \i \p s_n
	\notag \\[1ex] & \mspace{36.mu}
	= \ln \biggl(\frac{\sh(\h + \la - x_n)}{\sh(\h - \la + x_n)}\biggr)
        \int_{{\cal C}_n} \frac{\rd \m}{2 \p \i} \: \6_\m \ln \bigl(1 + \fa_n (\m)\bigr)
	\notag \\ & \mspace{72.mu}
        + \int_{{\cal C}_n} \frac{\rd \m}{2 \p \i} \:
	  \bigl(\cth(\la - \m + \h) - \cth(\la - \m - \h)\bigr) \ln (1 + \fa_n) (\m)
	\notag \\[1ex] & \mspace{36.mu}
	= - s_n \ln \biggl(\frac{\sh(\h + \la - x_n)}{\sh(\h - \la + x_n)}\biggr)
          - \int_{{\cal C}_n} \frac{\rd \m}{2 \p \i} \: K(\la - \m)
	       \ln (1 + \fa_n) (\m) \epp
\end{align}
Here we have to supply several comments and explanations. Generally,
some care is necessary, when we take the logarithm of a meromorphic
function and even more if we integrate it up along a contour. The
first logarithm under the integral on the left hand side is defined
by its principal branch. Then, due to our prerequisites, it defines
a holomorphic function of $\m$ inside and on the contour ${\cal C}_n$
for all $\la \in {\cal C}_n$. The logarithmic derivative under the
contour is meromorphic with simple poles with residue 1 at the
Bethe roots and a simple pole with residue $-N/2$ at $- \frac{h_R}{NT}$.
This explains the first equation. The second equation may be understood
as fixing the branch in the definition of the logarithm of the function
$\fa_n (\la)$. In the third equation we perform a partial integration
of the integral on the left hand side of the equation. This requires
that we define the function $\ln(1 + \fa_n)$ as a holomorphic function,
having no jumps of $2 \p \i$, as we move along the contour ${\cal C}_n$.
For this purpose we fix any point $x_n \in {\cal C}_n$ and define
a contour ${\cal C}_{x_n}^\la$ running from $x_n$ to $\la$ in positive
direction along ${\cal C}_n$. Then
\begin{equation}
     \ln(1 + \fa_n) (\la) =
        \int_{{\cal C}_{x_n}^\la} \rd \m \: \6_\m \ln(1 + \fa_n (\m))
	+ \ln \bigl(1 + \fa_n (x_n)\bigr) \epc
\end{equation}
where the rightmost logarithm is defined by its principal
branch, has the required properties.

Equation (\ref{derivenlie}) can be interpreted as nonlinear integral
equation for the auxiliary function $\fa_n$. For this purpose we rewrite
it in the form
\begin{multline} \label{nliefinite}
     \ln \bigl(\fa_n (\la)\bigr) = \\
	  - s_n \ln \biggl(\frac{\sh(\la - x_n + \h)}{\sh(\la - x_n - \h)}\biggr)
          - \frac{\e_0^{(N)} (\la)}{T}
          - \int_{{\cal C}_n} \frac{\rd \m}{2 \p \i} \: K(\la - \m)
	       \ln (1 + \fa_n) (\m) \epp
\end{multline}
Note that the explicit dependence on $x_n$ vanishes for states
with $s_n = 0$. Another possibility to simplify the appearance
of equation (\ref{nliefinite}) occurs if the contour ${\cal C}_n$
can be deformed in such a way that we can send $\Re x_n
\rightarrow - \infty$. Then
\begin{equation} \label{nlie}
     \ln \bigl(\fa_n (\la)\bigr) =
	  2 \i \g s_n - \frac{\e_0^{(N)} (\la)}{T}
          - \int_{{\cal C}_n} \frac{\rd \m}{2 \p \i} \: K(\la - \m)
	       \ln (1 + \fa_n) (\m) \epp
\end{equation}
\subsection{Back to the roots}
We just argued that every solution $\{\la_j^{(n)}\}_{j=1}^{M_n}$ of
the Bethe Ansatz equations corresponds to an auxiliary function $\fa_n$
subject to the monodromy condition (\ref{monocondition}) and solving
the nonlinear integral equation (\ref{nliefinite}). This reasoning
can be reversed, performing the following steps.
\begin{enumerate}
\item
Exponentiate (\ref{nliefinite}). It follows that $\fa_n$ has
an $N/2$-fold pole at $\la = - \frac{h_R}{NT}$ (located
inside~${\cal C}_n$).
\item
The monodromy condition (\ref{monocondition}) then implies that
$1 + \fa_n$ has precisely $M_n = N/2 - s_n$ roots inside ${\cal C}_n$.
\item
Going backwards through partial integration we see that these roots
must satisfy the Bethe Ansatz equations (\ref{bae}).
\end{enumerate}
Summing up, we have seen that the Bethe Ansatz equations are in
one-to-one correspondence to pairs $s_n, {\cal C}_n$, where the
${\cal C}_n$ denote equivalence classes of simple closed contours.
\subsection{Comments}
\begin{enumerate}
\item
We did not explain the algebraic Bethe Ansatz in any detail,
since several pedagogical accounts are available in the literature
(e.g.\ \cite{Thebook,Faddeev95,KBIBo}) and since it was covered by the
lectures of Nikita Slavnov during the Les Houches 2018 summer
school on Integrability in Atomic and Condensed Matter Physics.
\item
For the description of the thermodynamics of the spin chain
only the dominant eigenvalue $\La_0$ and the corresponding
auxiliary function $\fa_0$ are needed. These will be identified
in the next section.
\item
There are other methods for deriving non-linear integral equations
and representations of the eigenvalues involving their solutions,
most importantly the `method of functional equations'. For further
reading we recommend \cite{Suzuki99}.
\end{enumerate}

\section{Identification of the dominant state and free energy per
lattice site of the XXZ chain}
\subsection{Identification of the dominant state}
In order to describe the thermodynamics of the XXZ chain we
have to find out which spin value $s_0$ and which contour
${\cal C}_0$ belong to the dominant state. The answer can
be guessed by considering the special cases $\D = 0$,
$T \rightarrow 0$, $T \rightarrow + \infty$ and by performing
numerical calculations for small Trotter numbers $N$. For space-time
limitations we restrict ourselves to the consideration of the
high-$T$ limit here in which we obtain a particularly clear
and simple picture.
\begin{figure}
\begin{center}
\includegraphics[width=.80\textwidth]{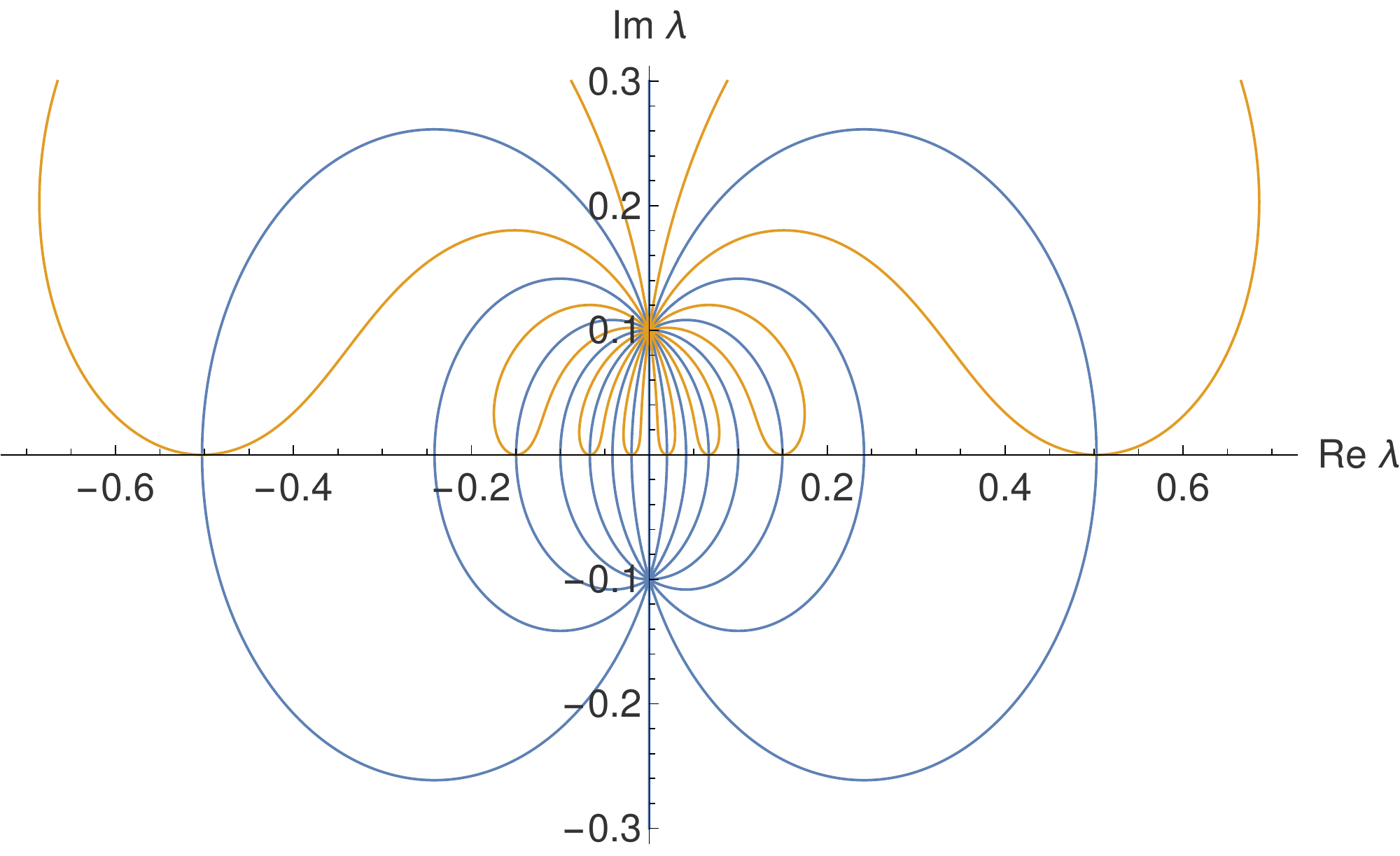}
\caption{\label{fig:feps} Contour plot of the function
$f_\e$ for $N = 16$, $\e = 0.1$. Blue contours $\Im f_\e = 0$,
orange contours $\Re f_\e = - 1$. We see $8 = N/2$ points
$\la_j$ on the real line at which $f_\e (\la_j) = - 1$.}
\end{center}
\end{figure}

\subsubsection{Bethe Ansatz at high temperature}
Let us have a closer look at the explicit form of the auxiliary
functions $\fa_n$ introduced in (\ref{defaux}). Recalling that
$h_R = - 2 \i J \sin (\g)$ for the XXZ chain and that we agreed
upon restricting $\g$ to the interval $(0,\p/2]$ for simplicity,
we see that
\begin{equation}
     \e = \frac{2J \sin(\g)}{NT} > 0
\end{equation}
and that this number becomes arbitrarily small for large $T$.
Inserting it into the explicit expression for $\fa_n$ we obtain
\begin{equation} \label{auxexpl}
     \fa_n (\la) = \re^{- \k/T}
                   \biggl(\frac{\sh(\la + \i \e)}
		               {\sh(\la - \i \e)}\biggr)^\frac{N}{2}
                   \biggl(\frac{\sh(\la + \i (\g - \e))}
		               {\sh(\la - \i (\g - \e))}\biggr)^\frac{N}{2}
                   \prod_{j=1}^{M_n}
		      \frac{\sh(\la - \la_j^{(n)} - \i \g)}
		           {\sh(\la - \la_j^{(n)} + \i \g)} \epc
\end{equation}
where $\{\la\}_n = \{\la_j^{(n)}\}_{j=1}^{M_n}$ is a solution
to the Bethe Ansatz equations $\fa_n (\la_j^{(n)}) = - 1$,
$j = 1, \dots, M$. We would like to find `perturbative solutions'
for small $\e > 0$. One particular such solution is almost
obvious. We shall see that it describes the dominant state
in the high-temperature limit. This solution is `generated'
by the second factor on the right hand side of (\ref{auxexpl}).
The latter has an $N/2$-fold pole at $\i \e$ and an $N/2$-fold
zero at $- \i \e$. If $\e$ is small, pole and zero are very
close to each other and a model of this function for small
$|\la|$ is
\begin{equation}
     f_\e (\la) = \biggl(\frac{\la + \i \e} {\la - \i \e}\biggr)^\frac{N}{2} \epp
\end{equation}
Close to the zero and close to the pole there are $N/2$ directions
in which the phase of this function is $\i \p$. They are connected
by lines on which $f_\e$ is real negative and goes from zero
to minus infinity. Since $f_\e (\la)$ takes values on the unit circle
for $\la$ on the real line, there are $N/2$ solutions $\la_j$, $j = 1,
\dots, N/2$, of the equation $f_\e (\la) = - 1$ on the real axis
which all go to zero for $\e \rightarrow 0$. This is sketched in
figure~\ref{fig:feps}. Thus, setting $M_n = N/2$ and inserting the
$\la_j$ for $\la_j^{(n)}$ into the fourth factor on the right hand
side of (\ref{auxexpl}), we see that the product of third and
fourth factor goes to $1$ for $\e \rightarrow 0$. Since the first
factor goes to $1$ as well in the high-$T$ limit, we have obtained
a special high-temperature solution.

\subsubsection{A special high-temperature solution}
In order to formalise this we set
\begin{equation} \label{larescale}
     \la_j = \frac{x_j}{T} \epc
\end{equation}
$j = 1, \dots, N/2$. We will look for a high-$T$ solution of the
Bethe equations with $|x_j| < R$ for some $R > 0$. Setting $\la = x/T$,
inserting (\ref{larescale}) into (\ref{auxexpl}) and sending
$T \rightarrow + \infty$ the Bethe Ansatz equations turn into
\begin{equation}
     \biggl(\frac{x - \frac{h_R}{N}}{x + \frac{h_R}{N}}\biggr)^\frac{N}{2}
        = - 1 \epc
\end{equation}
or, equivalently,
\begin{equation} \label{defphigh}
     p(x) = \Bigl(x - \frac{h_R}{N}\Bigr)^\frac{N}{2}
            + \Bigl(x + \frac{h_R}{N}\Bigr)^\frac{N}{2} = 0 \epp
\end{equation}
Now $p$ is a polynomial of order $N/2$ with asymptotics
$p(x) \sim 2x^{N/2}$ for $x \rightarrow \infty$. Thus, there
are $x_1, \dots, x_{N/2} \in {\mathbb C}$ such that
\begin{equation} \label{phighfactor}
     p(x) = 2 \prod_{j=1}^{N/2} (x - x_j) \epp
\end{equation}
\subsubsection{The corresponding eigenvalue}
The corresponding eigenvalue is
\begin{multline} \label{evainfinitt}
     \La (\la) = \re^\frac{\k}{2T} \prod_{j=1}^{N/2}
                 \frac{\sh(\la + \frac{h_R}{NT}) \sh(\la - \la_j - \h)}
		      {\sh(\la - \la_j) \sh(\la + \frac{h_R}{NT} - \h)}
               + \re^{- \frac{\k}{2T}} \prod_{j=1}^{N/2}
                 \frac{\sh(\la - \frac{h_R}{NT}) \sh(\la - \la_j + \h)}
		      {\sh(\la - \la_j) \sh(\la - \frac{h_R}{NT} + \h)} \\
                \xrightarrow[\scriptscriptstyle T \rightarrow + \infty]{\mspace{72.mu}} \qd
		\prod_{j=1}^{N/2} \frac{x + \frac{h_R}{N}}{x - x_j} +
		\prod_{j=1}^{N/2} \frac{x - \frac{h_R}{N}}{x - x_j} = 2 \epp
\end{multline}
\enlargethispage{-6ex}
Here we have used (\ref{defphigh}) and (\ref{phighfactor}) in the last
equation.
\subsubsection{Full spectrum the in high-temperature limit}
Observe that
\begin{multline}
     t_\infty = \lim_{T \rightarrow + \infty} t(0|\a) =
        \lim_{T \rightarrow + \infty} \:
        \text{\raisebox{-42pt}{\includegraphics[width=.52\textwidth]{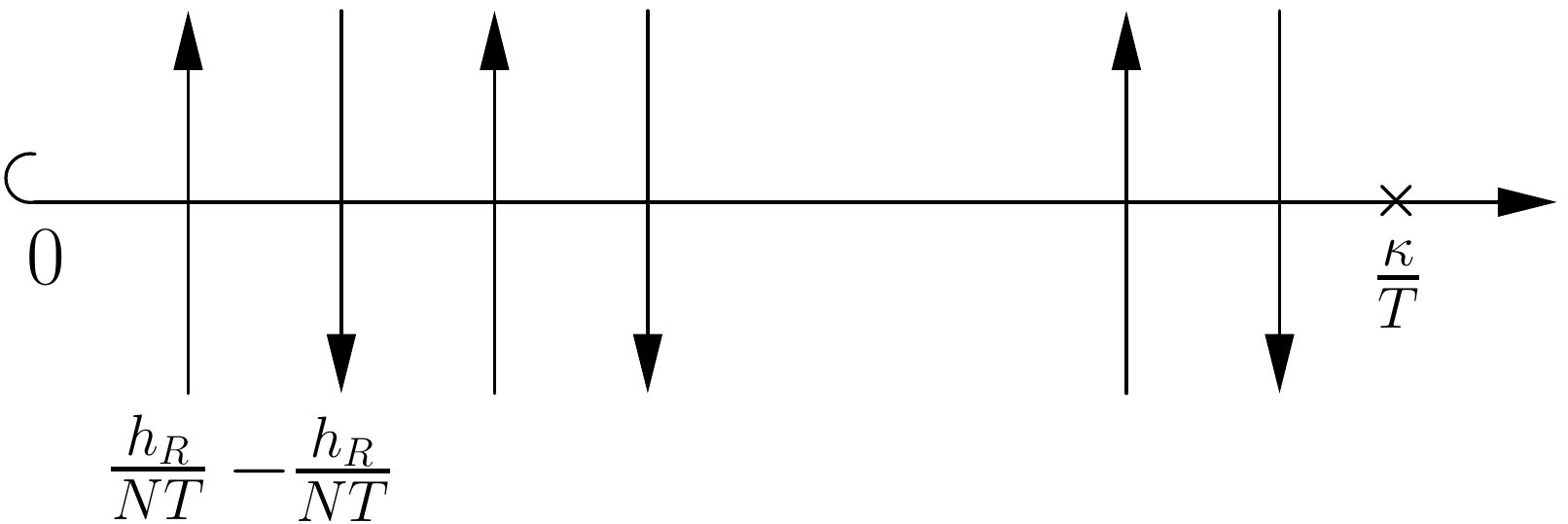}}}
     \\[1ex] = \:
        \text{\raisebox{-32pt}{\includegraphics[width=.52\textwidth]{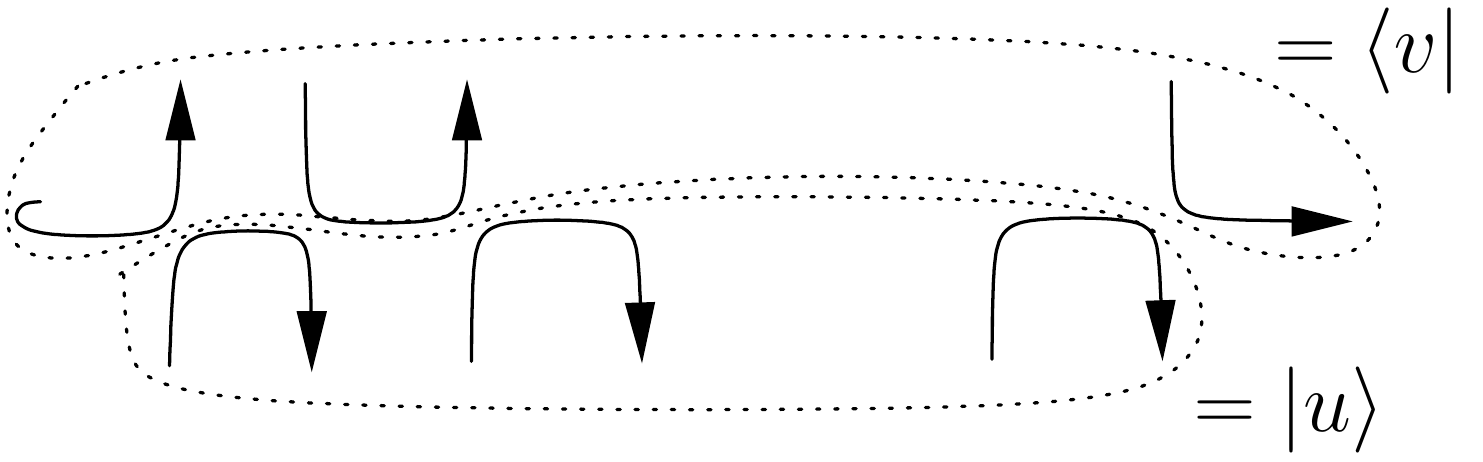}}}
        = \: |u\>\<v|
\end{multline}
due to the regularity (\ref{graphreg}) of the $R$-matrix. Clearly 
$|u\>\<v|$ is a one-dimensional projector. Moreover,
\begin{equation}
     \<v|u\> = \:
        \text{\raisebox{-25pt}{\includegraphics[width=.13\textwidth]{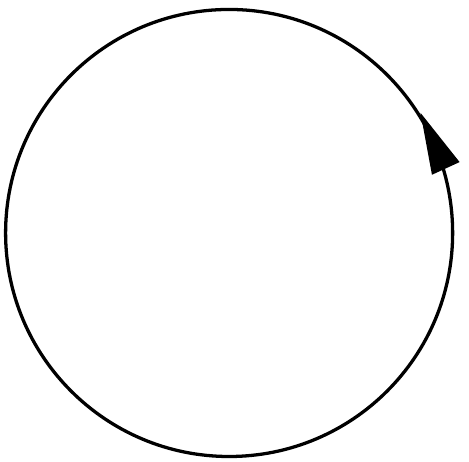}}}
	\: = 2 \epp
\end{equation}
Thus, the spectrum of $t_\infty$ is $\{2, 0\}$, where the eigenvalue
$0$ is $2^N - 1$-fold degenerate. This means that the dominant state
in the high-temperature limit is non-degenerate and has eigenvalue the
$\La (0) = 2$. Comparing with (\ref{evainfinitt}) above we see that
the corresponding Bethe roots are $\la_j = x_j/T$, $j = 1, \dots, N/2$,
where the $x_j$ are the roots of the polynomial $p$.
\subsubsection{Bethe roots of the dominant state in the Trotter limit}
\label{hightroots}
It is not difficult to calculate these roots explicitly. For this
purpose we have to solve the Bethe Ansatz equations in the
high-temperature limit,
\begin{equation} \label{baetinf}
     \biggl(\frac{x_j - \frac{h_R}{N}}{x_j + \frac{h_R}{N}}\biggr)^\frac{N}{2}
        = - 1 \epc
\end{equation}
$j = 1, \dots, N/2$. Clearly, if $x_j$ is a root, then $- x_j$ is
a root, and if $N/2$ is odd, then $x_j = 0$ is a root. Taking
the logarithm of (\ref{baetinf}) and setting
\begin{equation} \label{defphij}
     \tg \Bigl(\frac{\ph_j}{2}\Bigr) = \frac{2J \sin(\g)}{N x_j}
\end{equation}
we obtain, for any non-zero root $x_j$,
\begin{equation}
     \frac{N}{2} \ln \Biggl( \frac{1 + \frac{\i 2J \sin(\g)}{N x_j}}
                                  {1 - \frac{\i 2J \sin(\g)}{N x_j}} \Biggr)
        = \frac{\i N \ph_j}{2} = \i (2j - 1) \p \epp
\end{equation}
Using once more (\ref{defphij}) and solving for $x_j$ we
arrive at
\begin{equation}
     x_j = \frac{2 J \sin(\g)}{N \tg \bigl(\frac{(2j - 1)\p}{N}\bigr)} \epc
\end{equation}
where, due to the $\p$-periodicity of the tangent function, we may
restrict the range of $j$ to $- N/4 + 1 \le j \le N/4$ if $N/2$ is even or
$- N/4 + 3/2 \le j \le N/4 - 1/2$ if $N/2$ is odd.
\begin{figure}
\begin{center}
\includegraphics[width=.80\textwidth]{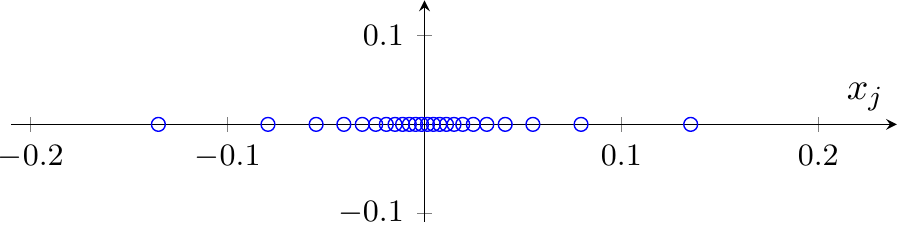}
\caption{\label{fig:xroots_example} Example for a configuration of
the roots $x_j$ for $N = 52$, $J = 1$, $\g = 0.7$.}
\end{center}
\end{figure}
In the Trotter limit, $N \rightarrow + \infty$, the roots get confined
in the interval $(2 J \sin(\g)/\p) \times [-1, 1]$ and accumulate
at the origin. The outer roots converge to $\frac{2J\sin(\g)}{(2j - 1) \p}$.
This behaviour is illustrated with an example in
figure~\ref{fig:xroots_example}.

Exercise: Find the other roots of the equation
\begin{equation}
     \fa (\la) = \re^{- \k/T}
        \prod_{j=1}^{N/2}
	\frac{\sh(\la - \frac{h_R}{NT})}{\sh(\la + \frac{h_R}{NT})}
	\frac{\sh(\la + \frac{h_R}{NT} - \h)}{\sh(\la - \frac{h_R}{NT} + \h)}
	\frac{\sh(\la - x_j/T + \h)} {\sh(\la - x_j/T - \h)} = - 1 \epp
\end{equation}
Answer: For $T \rightarrow + \infty$ we have $N/2$ roots close
to $+ \h$ and $N/2$ roots close to $- \h$, which can be seen by
setting $\la = z/T \pm \h$ and sending $T \rightarrow + \infty$.
\subsubsection{Dominant state contour}
For the dominant state we may thus choose the contour
\begin{multline} \label{hightczero}
     {\cal C}_0 = [- R - \i \g_-/2, R - \i \g_-/2] \cup
                  [R - \i \g_-/2, R + \i \g_-/2] \\ \cup
                  [R + \i \g_-/2, - R + \i \g_-/2] \cup
		  [- R + \i \g_-/2, - R - \i \g_-/2]
\end{multline}
with $R > 0$ large enough. As required, ${\cal C}_0$ encloses
all Bethe roots of the dominant state in the high-temperature
limit, but no other root of the equation $\fa_0 (\la) = - 1$
and no pole of this function other than the pole at
$\la = - \frac{h_R}{NT}$. Since $s = 0$ for the dominant
state, as seen above, we conclude that the auxiliary function of
the dominant state satisfies the nonlinear integral equation
\begin{equation} \label{nliedomi}
     \ln \bigl(\fa_0 (\la)\bigr) =
	  - \frac{\e_0^{(N)} (\la)}{T}
          - \int_{{\cal C}_0} \frac{\rd \m}{2 \p \i} \: K(\la - \m)
	       \ln (1 + \fa_0) (\m)
\end{equation}
with ${\cal C}_0$ according to (\ref{hightczero}).
\subsection{Trotter limit and free energy per lattice site}
\subsubsection{Trotter limit}
As follows from \ref{hightroots} the Bethe roots stay confined
close to $\la = 0$ for $N \rightarrow + \infty$. We therefore
obtain the auxiliary function in the Trotter limit by replacing
$\e_0^{(N)}$ by its limit
\begin{equation}
     \e_0 (\la) = \lim_{N \rightarrow \infty} \e_0^{(N)}
        = \k - 2 \i J \sin(\g) \re(\la) \epc \qd
	\re(\la) = \cth(\la) - \cth(\la + \h)
\end{equation}
in the nonlinear integral equation (\ref{nliedomi}), resulting in
\begin{equation} \label{nliedomitl}
     \ln \bigl(\fa_0 (\la)\bigr) =
	  - \frac{\e_0 (\la)}{T}
          - \int_{{\cal C}_0} \frac{\rd \m}{2 \p \i} \: K(\la - \m)
	       \ln (1 + \fa_0) (\m) \epp
\end{equation}
\subsubsection{Free energy per lattice site}
Consider the integral
\begin{align} \label{domieva}
     & \frac{\k}{2T} + \int_{{\cal C}_0} \frac{\rd \m}{2 \p \i} \:
                        \re(\m - \la) \ln (1 + \fa_0) (\m) \notag \\[1ex]
        & \mspace{18.mu} =
	  \ln \bigl(1 + \fa_0 (\la)\bigr) + \frac{\k}{2T}
	  \notag \\ & \mspace{54.mu}
	  + \int_{{\cal C}_0'} \frac{\rd \m}{2 \p \i} \:
	     \biggl(\6_\m \ln\biggl(\frac{\sh(\m - \la)}
	                                 {\sh(\m - \la + \h)}\biggr)\biggr)
	     \ln (1 + \fa_0) (\m) \mod 2 \p \i \notag \\[1ex]
        & \mspace{18.mu} =
	  \ln \bigl(1 + \fa_0 (\la)\bigr) + \frac{\k}{2T}
	  \notag \\ & \mspace{54.mu}
	  - \int_{{\cal C}_0'} \frac{\rd \m}{2 \p \i} \:
	     \ln\biggl(\frac{\sh(\m - \la)}{\sh(\m - \la + \h)}\biggr)
	     \6_\m \ln (1 + \fa_0) (\m) \mod 2 \p \i \notag \\[1ex]
        & \mspace{18.mu} =
	  \ln \bigl(1 + \fa_0 (\la)\bigr) + \frac{\k}{2T}
          + \ln\biggl(\frac{Q_0 (\la - \h)}{Q_0 (\la)}\biggr)
	  + \frac{N}{2} \ln\biggl(
	    \frac{\sh(\la + \frac{h_R}{NT})}{\sh(\la + \frac{h_R}{NT} - \h)}\biggr)
	    \mod 2 \p \i \notag \\[1ex]
        & \mspace{18.mu} =
	  \ln \bigl(\La_0 (\la)\bigr) \mod 2 \p \i \epp
\end{align}
Here ${\cal C}_0'$ is a modification of the contour ${\cal C}_0$
such that ${\cal C}_0 - {\cal C}_0'$ is a small positively oriented
circle around $\la$. In the partial integration in the second 
equation we have used that $s = 0$, implying that there are no
boundary terms. Equation (\ref{domieva}) determines the eigenvalue
in the Trotter limit.

Recalling (\ref{ftl}) we obtain the free energy per lattice site
of the XXZ chain in the thermodynamic limit,
\begin{equation} \label{fintrep}
     f(T, h) = - \frac{\k}{2} - T \int_{{\cal C}_0} \frac{\rd \la}{2 \p \i} \:
                                     \re(\la) \ln (1 + \fa_0) (\la)
\end{equation}
where $\fa_0$ is the solution of the nonlinear integral equation
(\ref{nliedomitl}).

For the identification of the dominant state and the corresponding
auxiliary function we have considered the high-temperature limit
here. This brought us to the conclusion that $s_0 = 0$ and that
a possible contour ${\cal C}_0$ is the contour defined in (\ref{hightczero}).
There are many good reasons to believe that (\ref{fintrep}) and
(\ref{nliedomitl}) with the same choice of the contour hold for
all $T > 0$.
\subsection{Comments}
\begin{enumerate}
\item
As we mentioned in the introduction, the latter claim is supported by
numerical studies at finite Trotter number (for a pedagogical
review see \cite{GoSu10}), by a low-temperature analysis (see
e.g.\ \cite{DGK13a}) and by considering the XX chain (this is
recommended as an exercise, for some information see \cite{GKKKS17}).
In addition we would like to recommend the work \cite{GGKS18pp},
where the case of high but finite temperature was treated with
full mathematical rigour.
\item
The high-temperature analysis presented for the dominant state can
be extended to obtain a large class of excited states in the
high-temperature limit. We shall only sketch the calculation and
leave the details as an exercise. Let us look for a solution
$\{\la_j\}_{j=1}^M$ of the Bethe Ansatz equations (\ref{bae}) that
has the following high-temperature asymptotics:
\begin{subequations}
\begin{align}
     & \lim_{T \rightarrow + \infty} \la_j \ne 0 \qd
       \text{for $j = 1, \dots, n$} \epc \\
     & \la_j \sim \frac{x_j}{T} \qd
       \text{with $|x_j| < R$ for some $R > 0$ for $j = n + 1, \dots, M$.}
\end{align}
\end{subequations}
Inserting this high-temperature Ansatz into the Bethe Ansatz equation
(\ref{bae}) and performing the limit $T \rightarrow + \infty$, we see
that the first $n$ equations decouple and become
\begin{equation} \label{higherlevel}
     \biggl(\frac{\sh(\la_j + \i \g)}{\sh(\la_j - \i \g)}\biggr)^{s + n}
        \prod_{k=1}^n \frac{\sh(\la_j - \la_k - \i \g)}{\sh(\la_j - \la_k + \i \g)} = - 1 \epc
\end{equation}
for $j = 1, \dots, n$. These equations can be interpreted as a set
of so-called higher-level equations for the high-$T$ limit. They
resemble the Bethe Ansatz equations of the spin-$1$ XXZ chain.

Taking the product over all $j = 1, \dots, n$ in (\ref{higherlevel})
we obtain the `momentum quantisation condition'
\begin{equation} \label{momquant}
     \biggl(\prod_{j=1}^n \frac{\sh(\la_j + \i \g)}
                               {\sh(\la_j - \i \g)}\biggr)^{s + n} = 1
        \qd \Leftrightarrow \qd
        \prod_{j=1}^n \frac{\sh(\la_j + \i \g)}
	                   {\sh(\la_j - \i \g)} = \re^\frac{2 \p \i \ell}{s+n}
\end{equation}
for some $\ell \in \{0, 1, \dots, s + n - 1\}$ that depends on
$\{\la_j\}_{j=1}^n$. Inserting the $\la_j$, $j = n+1, \dots, M$, into
the Bethe Ansatz equations (\ref{bae}), performing the limit
$T \rightarrow + \infty$ and using (\ref{momquant}), we obtain a
set of equations that determine the $x_j$,
\begin{equation} \label{baetexcinf}
     \biggl(\frac{x_j - \frac{h_R}{N}}{x_j + \frac{h_R}{N}}\biggr)^\frac{N}{2}
        = (-1)^{s + n - 1} \re^{- \frac{2\p\i\ell}{s+n}} \epp
\end{equation}
Depending on $N, \ell, s, n$ this equation may admit a root $x_j = 0$.
All other roots are given by
\begin{equation}
     x_j = \frac{2 J \sin(\g)}
                {N \tg \bigl(\frac{2\p}{N}\bigl(k(j) - \frac{\ell}{s + n}\bigr)\bigr)} \epc
\end{equation}
where $k(j)$ is integer, if $s + n$ is even, or half-odd integer, if
$s + n$ is odd. This means that we may choose the $M - n$ roots $x_j$
from a set of $N/2$ inequivalent values, giving $\binom{N/2}{M - n}$
different solutions.

In \cite{GGKS18pp} the high-$T$ limit was worked out on more
rigorous grounds, starting from the nonlinear integral equations
for the excited states.
\end{enumerate}

%\clearpage

\bibliographystyle{SciPost_bibstyle}
\bibliography{hub}

\end{document}